\documentclass[twocolumn]{emulateapj}
\shortauthors{Subrahmanyan, Saripalli, Safouris and Hunstead}
\shorttitle{Giant radio galaxy and large-scale structure}
\begin{document}

\title{On the relationship between a giant radio
  galaxy MSH~05$-$2{\it 2} and the ambient large-scale galaxy structure}

\author{Ravi~Subrahmanyan,\altaffilmark{1,2} 
Lakshmi~Saripalli,\altaffilmark{1,2} Vicky Safouris,\altaffilmark{2,3} 
and Richard~W.~Hunstead\altaffilmark{4} }

\altaffiltext{1}{Raman Research Institute, C V Raman Avenue, Sadashivanagar, 
Bangalore 560 080, India.}
\altaffiltext{2}{Australia Telescope National Facility, CSIRO, 
Locked bag 194, Narrabri, NSW 2390, Australia.}
\altaffiltext{3}{Research School of Astronomy and Astrophysics, Mount Stromlo
Observatory, Australian National University, Cotter Road, 
Weston ACT 2611, Australia.}
\altaffiltext{4}{School of Physics, University of Sydney, NSW 2006, Australia.}

\begin{abstract}
We present a comparison of the properties of a giant radio galaxy and
the ambient intergalactic medium, whose properties are inferred from the
large-scale distribution in galaxies.  The double 
lobes of the radio galaxy MSH~05$-$2{\it 2} are giant---1.8~Mpc projected
linear size---and interacting with the environment outside the 
interstellar medium and coronal halo associated with the host 
galaxy. The radio lobes appear to be relicts and the 
double structure is asymmetric. We have examined the 
large-scale structure in the galaxy distribution surrounding the 
radio source. The host galaxy of MSH~05$-$2{\it 2} is associated with 
a small group that lies close to the boundary of sheet-like and 
filamentary density enhancements, and adjacent to a void. 
Assuming that the galaxies trace gas, the asymmetries in the radio
morphology in this case study appear related to the anisotropy in the medium.
However, the observed overdensities and structure formation models for 
the heating of the intergalactic medium (IGM)   
suggest a density-temperature product for the IGM
environment that is an order of magnitude below that expected from the
properties of the radio source. 
The discordance suggests that even sources like MSH~05$-$2{\it 2}, which are
observed in the relatively low-density IGM environment associated with the
filamentary large-scale structure and
have multiple signatures of being relicts, may be
overpressured and evolving towards an equilibrium relaxed state 
with the ambient IGM. In this picture, relaxed relicts outside cluster and
group environments would have surface brightness well below that observed in
MSH~05$-$2{\it 2} and the detection 
limits of current wide-field radio surveys. 
Alternately, it is speculated that astrophysical feedback originating in 
galaxy overdensities observed 1--2~Mpc to the N and
NE of MSH~05$-$2{\it 2} might be the mechanism for the heating of
the ambient IGM gas. In such a scenario, MSH~05$-$2{\it 2} would be a relaxed
relict in equilibrium with the IGM environment and such feedback might
additionally aid in the formation of 
the observed asymmetries in the radio morphology.
\end{abstract}


\section{Introduction}

Young radio sources evolve within the host interstellar medium; as 
they grow larger the radio structures presumably interact with the 
coronal X-ray halos \citep{Fo85, Su04} of their host elliptical 
galaxy. Not only the lobe morphology, but also the lobe 
separation asymmetries, are 
believed to be influenced by the environment in which the radio 
galaxy evolves. For example, associated extended emission line 
regions have been shown to retard the growth of the lobe on that 
side \citep{Mc91}.

Galaxies that host active galactic nuclei (AGNs) occur in a range of 
environments and the jets from the central engines in these galaxies 
encounter a variety of conditions as they emerge into the 
intergalactic medium (IGM).  Membership and proximity to clusters, 
groups, and the filaments and sheets in the large scale structure of 
the Universe might all determine the gaseous intergalactic 
environment and these would, additionally, depend on cosmic epoch.  
Within clusters of galaxies, the prevailing intracluster winds and 
galaxy peculiar velocities shape the lobes of double radio sources 
into head-tail and wide-angle morphologies \citep{Bl98}.  Outside 
clusters and groups, giant radio sources whose linear sizes are in 
the range of several megaparsecs evolve in the more tenuous 
intergalactic medium, the nature of which is obscure.  Asymmetries 
caused by jet-IGM interactions would be expected to be manifest 
particularly in circumstances where the host galaxies are at the 
edge of galaxy clusters or filamentary structures or more generally 
at the edge of galaxy overdensities.

\citet{Sa86} investigated the pronounced asymmetry in the lobe 
lengths of a small number of relatively nearby giant radio galaxies 
and found that the shorter lobe was preferentially located 
on the side that had excess galaxies over the background. The 
asymmetry was presumably a result of the higher density of gas 
associated with galaxy clustering seen on that side; this is 
consistent with the expectation that the large-scale intergalactic 
gaseous medium would be traced by the distribution of galaxies.

Large scale redshift surveys have revealed the richness in galaxy 
distribution structures---galaxies are observed to form 
criss-crossing chains and sheets several megaparsec wide with 
regions largely devoid of galaxies in between. Most giant radio 
galaxies do not belong to rich clusters of galaxies; they tend to 
populate regions with low galaxy density, corresponding to galaxy 
filamentary structures. Therefore, the medium associated with this 
component of the large scale structure is of relevance. While galaxy 
clusters are known to be permeated with hot thermal gas (seen via 
X-ray emission), little is known regarding gas associated with  
galaxy filaments.

Simulation studies of galaxy formation \citep{Ce99,Da01} have 
suggested the existence of a warm-hot intergalactic medium (WHIM) at 
low redshifts. The puzzling deficit in the baryon content of the 
low-redshift Universe was identified with the  WHIM: a gas that is 
intermediate in temperature between warm photo-ionized gas in voids and hot
intracluster gas, and is too low in density to be easily detectable in X-rays.
The bulk of this gas is believed to be distributed along galaxy 
filaments and traced by moderate overdensities; it mostly lies 
outside of virialized structures like groups or clusters. The gas is 
believed to be enriched via outflowing galactic superwinds from 
supernovae in galaxies; as a consequence, highly ionized oxygen and 
neon are considered important species that would render the WHIM 
visible via absorption of the continuum emission of background 
quasars. There are several ongoing searches for this absorption 
signature and today there is tantalizing evidence for the WHIM via 
such methods (see, for example, \citet{Ni05}). However, we recognise
that the background AGN population is sparse, and sparser 
still are powerful AGN capable of rendering robust absorption 
detections.  Additional and independent probes of this key component 
of the intergalactic medium are, therefore, highly desirable.

The WHIM gas is believed to have been heated as a result of 
hydrodynamic shocks during the non-linear evolution of the 
large-scale filamentary structure in the Universe. However, there 
are likely to be additional complex astrophysical feedback 
mechanisms that might very well be a significant contributor to the 
entropy content of the WHIM.  This is an additional motivation for 
observations that probe the WHIM gas properties and provide 
constraints on feedback mechanisms.

In this paper we have for the first time used radio observations of 
a 1.8~Mpc giant radio galaxy MSH~05$-$2{\it 2} 
(referred to as 0503$-$286 by \cite{Sa86} and \cite{Su86}) 
and galaxy redshift surveys to
compare the properties of the synchrotron plasma lobes with the 
properties of the ambient IGM, which is believed to be the WHIM in this
particular case. Motivated by  
the earlier hypothesis that the lobe extent asymmetry in this source 
might be related to the asymmetry in the sky distribution in galaxy 
number density, we have measured redshifts of galaxies over a large 
area around the radio source. Using these new data, along with 
archival redshift measurements, we have constructed the 3D 
distribution of galaxies around the radio source and identified 
galaxy filaments that might be responsible for the asymmetry. Our 
modeling of the evolution of this radio source leads to inferences 
for the properties of the gas associated with the large-scale 
structure. With this case study we illustrate how giant radio 
galaxies may be useful probes of the IGM gas.

We organise the paper as follows: we begin with an introduction to 
the giant radio galaxy MSH~05$-$2{\it 2} in Section~2. In Section~3 
we present our new Very Large Array (VLA) and Molonglo Observatory Synthesis
Telescope (MOST) images of this radio 
source. The next section describes the galaxy redshift data: first 
we discuss the optical field over nearly $45\degr$ radius and then 
we present our 2-degree field (2dF) spectroscopy with the 
Anglo-Australian telescope (AAT). In Section~5 we derive physical 
properties of the radio source. Section~6 presents an estimation of the
properties of the ambient intergalactic medium for comparison with the
expectations based on the internal properties of the source; section~7
examines plausible causes for asymmetries in the source; section~8 is a
concluding discussion.

Throughout we adopt a flat cosmology with Hubble constant $H_{o}$ = 
71~km~s$^{-1}$~Mpc$^{-1}$ and matter density parameter $\Omega_{m} = 
0.26$. The optical host for the radio source is the luminous elliptical galaxy
ESO~422-G028 (also cataloged as MCG~$-$05-13-003; with J2000.0 coordinates
R.A. 05$^{h}$ 05$^{m}$ $49\fs22$, decl. $-$28$\degr$ 
35$\arcmin$ $19\farcs4$).
The redshift of the giant radio galaxy was first measured by 
\citet{Sa86} and \citet{Su86}; more recently the 6dF redshift survey 
\citep{Jo04} gave an improved value of z=0.038286. At this redshift 
$1\arcmin=45$~kpc and MSH~05$-$2{\it 2}, with an angular size of 
$40\arcmin$, has a projected linear size of 1.8~Mpc.

\section{MSH~05$-$2{\it 2}}

MSH~05$-$2{\it 2} was discovered to be a giant radio galaxy by two 
quite independent methods \citep{Sa86,Su86} and discussed in 
\citet{Sa88}.  Arcmin-resolution radio images in the frequency range 
843~MHz to 2.7~GHz were presented there and show a 
large-angular-size double, with disjoint elongated lobes oriented 
roughly NS and with a core component lying in between at the 
location of the host galaxy. The double radio source is one of the 
most asymmetric among giant radio sources: as seen projected on the 
sky, the southern lobe of this radio galaxy extends twice as far 
from the core compared to the northern lobe. An additional 
morphological peculiarity is that while the core, southern jet and 
southern lobe appear collinear, the northern lobe appears 
significantly displaced to the west of this radio axis.  
\citet{Sa86} also noted the presence of a higher density of galaxies 
on the sky to the north and east of the host galaxy and suggested 
that the asymmetric galaxy distribution 
might be a clue to the cause of the morphological 
peculiarities. The radio observations presented in \citet{Sa86} and 
\citet{Su86} were followed up with VLA observations at 20 and 6~cm 
wavelengths in order to image the detailed structure and 
polarization in the lobes and the core. Preliminary images were 
presented in \citet{Sa88}; in this paper we present a re-analysis of 
the VLA data that has resulted in significantly better quality 
images in total intensity and polarization with much improved 
dynamic range.

\section{Radio continuum imaging}

An 843~MHz image of MSH~05$-$2{\it 2} made with the 
MOST is shown in Fig.~1.  The 
image has been made with a beam of full width at half maximum (FWHM) 
$94\farcs0 \times 45\farcs0$ 
at a position angle (P.A.) of $0\degr$ and has an rms noise of 
0.8~mJy~beam$^{-1}$. At 843~MHz, the flux densities of the northern 
and southern lobes are 2.82 and 2.04~Jy respectively, where the 
systematic calibration errors are less than 5\%.

MSH~05$-$2{\it 2} was observed in the CnB array configuration of the 
Very Large Array (VLA). A configuration with the long northern arm 
was selected so that projected baselines would have roughly equal 
ranges in NS and EW despite the southern declination of the source.  
Observations were made in 6-hr sessions around transit on 19 and 20 
October, 1986.  Observations in the 20-cm band were made using a 
pair of 25-MHz bands centered at 1415 and 1635~MHz. These 
observations were made in mosaic mode with two antenna pointings 
separated in declination by $25\arcmin$ to cover the two lobes of 
the NS oriented double radio source.  Observations in the 6-cm band 
were made using a pair of 50-MHz bands centered at 4835 \& 
4885~MHz; these observations were made using a single antenna 
pointing towards the radio core.  The schedule cycled through 20-cm 
observations towards the two lobes, 6-cm observations towards the 
core and calibrators.

The visibility data in the individual pointings and in each of the 
observing bands were separately calibrated in amplitude, phase and 
for instrumental polarization using standard AIPS procedures.  The 
flux density scale was set on the scale of \citet{Ba77} using 
observations of 3C48 at all frequencies.  Complex gain calibration 
was initially done using frequent observations of PMN~J0453$-$2807; 
the 20-cm data were then iteratively self-calibrated in 
amplitude and phase to improve the dynamic range.  The antenna 
polarization leakage corrections were derived from the data on 
PMN~J0453$-$2807, which covered a wide range of parallactic angles, 
and the phase angle between the orthogonal polarizations was 
determined using the observation of the linearly polarized source 
3C286.

The calibrated 20-cm visibility data for the different pointings 
were mosaic-imaged in MIRIAD and the Stokes I, Q \& U images were 
jointly deconvolved using the maximum entropy deconvolution 
algorithm implemented in PMOSMEM \citep{Sa99}.  The images have beam 
FWHM $15\farcs5 \times 14\farcs2$ at a P.A. of 8\fdg6 and rms noise 
of $40~\mu$Jy~beam$^{-1}$.  Images of the northern and southern 
lobes are shown separately for clarity; total intensity images are 
shown in Figs.~2 \& 3, and the distribution in polarized intensity 
and E-field orientation are shown in Figs.~4 \& 5. The total flux 
densities of the northern and southern lobes are, respectively, 1.55 
and 1.41~Jy, at 1520~MHz. The rotation measure (RM) towards the 
lobes was estimated by imaging separately the polarization at 1415 
and 1635~MHz; with a beam of FWHM $17\arcsec$, the RM was observed 
to be uniform over the entire source with a mean value of 
$+21$~rad~m$^{-2}$ and an rms scatter of 11~rad~m$^{-2}$. The 
polarization vectors in Figs.~4 \& 5 have been corrected for a 
uniform RM of $+21$~rad~m$^{-2}$.

The 6-cm data were not self-calibrated; the visibilities at both 
frequencies were imaged together in AIPS.  The radio image of the 
core region, made with a beam of FWHM $5\farcs0 \times 4\farcs1$ at 
a P.A. of $50\fdg8$ is in Fig.~6. The image has an rms noise of 
25~$\mu$Jy~beam$^{-1}$. The polarization vectors have been corrected 
assuming a uniform RM of $+21$~rad~m$^{-2}$.

\section{The large-scale galaxy distribution around MSH~05$-$2{\it 2}}

We extracted archival redshift measurements from the 6dF galaxy 
redshift survey \citep{Jo04} over a large sky area around 
MSH~05$-$2{\it 2}. The 6dF survey was complete to apparent magnitude limit
$b_{j} = 16.75$. Only galaxies with a redshift quality scale of 3 
(for probable redshifts) and 4 (for reliable redshifts) were 
retained, others being spectra for which redshifts could not be 
determined, tentative redshifts or stars. Fig.~7 shows the 6dF 
galaxy distribution within $45\degr$ of the host galaxy; the data 
are displayed in different redshift slices relative to the host 
galaxy redshift. We have adopted the gnomonic (tangent plane) 
projection for the representations and derive distances from 
redshifts neglecting peculiar velocities. As can be seen in panel 
(h), where all the 6dF survey galaxies in the sampled area are 
plotted together, the data available to date suffers from 
incompleteness and patchy coverage; nevertheless, patterns are 
clearly discernible in moving across the panels containing redshift 
slice data and the coverage is relatively complete close to the 
location of the host galaxy.

In Fig.~7 the six panels from (a) to (f) show the galaxy 
distribution in successive redshift slices 21~Mpc deep along the 
line of sight; panels (a) to (c) show slices in front of the host 
galaxy whereas panels (d) to (f) show slices beyond the host 
galaxy.  The panels are $80\degr$ wide corresponding to a linear 
size of 216~Mpc. The most prominent structure is a long 
filament-like feature exceeding 100~Mpc that is oriented in the 
NW-SE direction and crosses the sky just NE of the host galaxy. This 
feature is observed in panels (b) and (c) and does not appear in the 
other redshift slices; we infer the structure to be a sheet of 
galaxies at least 160~Mpc long that is oriented almost edge-on and 
extending nearly 40~Mpc from the host galaxy towards us. The 
projected width of the galaxy sheet is about 10~Mpc. In panel (g) we 
display all galaxies up to 42~Mpc in front of the host galaxy; this 
combines panels (b) and (c) and shows most clearly the NW-SE 
filament in the galaxy distribution close to MSH~05$-$2{\it 2}. Panel 
(g) also shows a sharp fall in the surface density of the galaxies 
at the centre on the SW side of the host. The galaxies in the sheet 
appear to be closer to the sky position of the host galaxy in panel 
(c) as compared to (b); the host is, therefore, inferred to be 
located on the farther and lower (SW) end of the galaxy sheet.  
Towards the northern end of the sheet, the distribution may be 
affected by incompleteness in the 6dF survey. In panels (d) and 
(e)---which display galaxies to distances of 21 and 42~Mpc beyond 
the host---there is a concentration of galaxies seen in the range 
3--$5\degr$ (8--13.5~Mpc) to the north of the host galaxy.

Using tools that enable visualization of the galaxy distribution in 
3D reveals additional richness: the sheet is in reality an edge-on 
complex network of curving filaments with a branch that runs close 
to and beyond the host galaxy connecting the condensation---that is 
seen in panel (d) to the north of the host---to the filaments 
composing the sheet seen in panels (b) and (c). All of this 
structure is located to the N and NE of the host, extending in 
redshift space on either side of the host, leaving the space to the 
SW of the host relatively devoid of galaxies.

The catalog of superclusters presented by \citet{Ka95} records an 
entry for a supercluster with an over-density of 20--40 at the 
location of the sheet/filament. The host elliptical galaxy is also 
noted as being in a cluster in the ESO/UPPSALA survey \citep{La82}. 
The linear feature in the galaxy distribution is also discernible in 
the panels in Fig.~4 of \citet{Jo04}, which depict the 6dF galaxy 
redshift survey.

\subsection{The 2dF  spectroscopy}

With the aim of examining the detailed structure of the galaxy 
distribution in the immediate vicinity of the host galaxy, we used 
the 2dF facility at the 3.9-m Anglo Australian Telescope (AAT) to 
make a redshift survey of galaxies within $1\degr$ of the radio 
source.  Up to 400 simultaneous spectra may be obtained towards sky 
positions within the $2\degr$-diameter field of the 2dF. We observed 
two lists of target galaxies, as separate allocations, using low 
resolution 300B gratings in the spectrographs; each target list was 
observed in three 20-min exposures.  The 2dF service observations 
were made on 2004 December 19 in occasional thin cloud and with a 
bright moon; the seeing was 1$\arcsec$--$1\farcs5$.

The master object list for the spectroscopy was created using a 
SuperCOSMOS selection of galaxies in a $2\degr$ diameter field 
centred at R.A. 05$^{h}$ 03$^{m}$ $51\fs0$, decl. $-$28$\degr$ 
45$\arcmin$ 00$\arcsec$ (B1950.0). This position is located to the 
south of the radio core and roughly equidistant from the two lobes. 
The extracted list contained all objects deemed to be galaxies, to a 
magnitude limit of $b_{j} = 20$. Nine SuperCOSMOS digitized sky 
survey images that were $15\arcmin$ wide---one centered at the field 
center with 8 flanking fields---were examined by eye to verify the 
robustness of the recognition of galaxies in the SuperCOSMOS
extraction. Most of the brighter galaxies appeared to have been 
selected correctly; those missed were at faint magnitudes with 
$b_{j} \ga 18.5$. 25 galaxies, including the host, that had 
redshifts measured in the 6dF survey were omitted from the 2dF 
observations.

The galaxy lists for the two 2dF fibre allocations were constructed 
using the CONFIGURE routine.  For this, the guide stars were 
assigned the highest priority of 9 and the galaxies had priorities 
assigned on the basis of their magnitude:  priority 8 for $15 < 
b_{j} < 17$, priority 7 for $17 < b_{j} < 18.5$ and priority 6 for 
$18.5 < b_{j} < 20$. The first list contained all galaxies with 
$b_{j} < 19$. To minimize cross-talk between the fibres, bright 
galaxies ($b_{j} < 17.5$) were separated from fainter galaxies by 
forcing their allocations to separate spectrographs. 
 %
 %
 The second list contained the 91 galaxies from the master list with 
$b_{j} < 19$, that remained unallocated in the first run of 
CONFIGURE, together with fainter galaxies with $19.0 < b_{j} < 
19.2$.  Seven galaxies among the unallocated objects had $b_{j} < 
17.5$; these were once again forced to the first spectrograph. The 
remaining galaxies in the unallocated list were assigned to the 
second spectrograph and galaxies with $19 < b_{j} < 19.2$ were 
allowed to be assigned by CONFIGURE to either of the two 
spectrographs. The data reduction was carried out using the 2DFDR 
software package; 1-dimensional spectra were derived after 
calibration. Excluding calibrations, there were in all 592 object 
spectra. The 6dF tool RUNZ was used to estimate redshifts; 
those with poor quality results were examined manually and redshifts 
were re-assigned in many cases.  Only redshifts of good quality 
were retained.

Including the 6dF galaxies, we have 359 galaxies with acceptable 
redshift measurements in the $2\degr$ diameter field around the 
radio source giving a pencil-beam view of the galaxy distribution 
close to MSH~05$-$2{\it 2}.  A histogram of the redshift distribution 
of galaxies within this $2\degr$ pencil beam is shown in Fig.~8 over 
the range $z=0.03$--0.05, where it is seen that the host galaxy, at 
$z=0.0383$, lies well within a tight clustering of galaxies in 
redshift space. The sky distribution of the 359 galaxies is shown in 
Fig.~9; large symbols mark galaxies that lie within $\pm 
0.003$ of the host galaxy redshift and small symbols show the 
remainder. 

At the redshift of the host, the $2\degr$ field has a diameter of 
5.4~Mpc and is centered $\approx 250$~kpc south of the host galaxy.  
The galaxies in Fig.~9 that fall in the narrow redshift slice 
containing the host galaxy are almost exclusively in the NE half of 
the $2\degr$ diameter circle.  
Among these galaxies, which are represented using large symbols in Fig.~9, the
concentration seen just NNE of the northern lobe is at somewhat higher
redshift relative to the host galaxy, whereas those galaxies represented by
large symbols and which have somewhat lower redshifts relative to the host
have a wider distribution over the semi-circle. 
The redshifts of the galaxy group at the location of the host have a 
distribution that spans the host redshift, indicating that the host 
is likely to be a member of this small group which is located at the 
fringe of a large-scale galaxy overdensity. The galaxy distribution 
in the 2dF pencil beam is consistent with the 6dF wide-field galaxy 
distribution and nicely provides a zoomed-in view of the immediate 
environment of MSH~05$-$2{\it 2}.

\section{Giant radio galaxy MSH~05$-$2{\it 2} revisited}

Although the radio lobes in MSH~05$-$2{\it 2} have an edge-brightened 
FR-{\sc ii} type structure, no compact hotspot-like features are 
observed associated with the lobes. In this context, it may be noted 
that the unresolved source observed within the northern lobe (see 
Fig.~2) has no intermediate structure suggesting any relationship to 
the radio lobe; it is probably an unrelated background source.

The total intensity distribution in both the lobes shows a wealth of 
structure. In the northern lobe a dip in intensity is observed close 
to the center and the brightness in the surrounding shell-like 
structure is higher by up to a factor four. In the southern lobe, 
the total intensity distribution is center brightened with 
surface-brightness fluctuations in a smaller range.

The radio images show that the two lobes, which are enveloped in a 
continuous bridge of low surface brightness emission, have the 
following overall asymmetries:
(a) the two lobes extend to very different distances from the core: 
the total extents of the source to the
north and south of the core are in the ratio 1:1.6;
(b) the backflow from the southern lobe extends three-quarters of 
the 
way towards the radio core, whereas the backflow from the northern 
lobe not only appears directed about $20\degr$ to the west of the 
core but also extends beyond the core; and
(c) the twin jets from the core are asymmetric: the brighter jet is 
to the south and directed towards the peak at the end of the 
southern lobe.

The centroid of MSH~05$-$2{\it 2} is located $2\farcm4$ 
west and $2\farcm7$   
south of the core, amounting to a projected offset of about 
$3\farcm6$ or 160~kpc to the SW.

From Figures~4 and 5 the two lobes of the giant radio galaxy are 
seen to be highly polarized.  The fractional polarization is about 
25--30\% in patches in the interior of the two lobes. Surrounding 
these patches the fractional polarization dips to below 10\% in 
ring-like structures and the fraction rises to above 50\% in the 
rims of both lobes where the typical circumferential projected 
magnetic field structure is seen.

The host galaxy has been cataloged as a ROSAT X-ray source, 
1WGA~J0505.8$-$2835, and we estimate its X-ray luminosity to be $7 
\times 10^{33}$~J~s$^{-1}$ in the 0.2--2~keV band. The $V$-band 
absolute optical magnitude of the host galaxy is estimated to be 
$M_{V}=-21.53$ which is typical of host galaxy magnitudes of 
powerful radio galaxies. The total radio power at 1.4~GHz is $1.07 
\times 10^{25}$~W~Hz$^{-1}$ which---considering the optical 
magnitude of the host galaxy---places the source on the dividing 
line between FR~{\sc i} and FR~{\sc ii} type radio sources. The 
total power is relatively low compared to double radio sources 
with edge-brightened structure.

Twin jets have been detected from the core. The jet, counterjet and 
core as well as the southern lobe are collinear. However, the 
northern jet does not appear to be directed towards the northern 
lobe.

The radio core, which is coincident with the host galaxy, has a flux 
density of $6.2 \pm 0.2$~mJy at 4860~MHz and $10.0 \pm 0.5$~mJy at 
1520~MHz.  Using images with a beam FWHM of $15\farcs5 \times 
14\farcs2$, the spectral index $\alpha$ between 1520 and 4860~MHz is 
$-0.3 \pm 0.1$ for the core component, $-0.4 \pm 0.1$ in the 
southern jet and $-0.5 \pm 0.1$ in the northern counterjet (the 
spectral index $\alpha$ is defined using the relation $S_{\nu} 
\propto \nu^{\alpha}$). The jet (Fig.~6) is detected to the south to 
a projected distance of 90~kpc and along this distance the B-field 
appears directed along the jet axis.  The jet is also knotty in 
appearance and has a constant opening angle of about $10\degr$ and 
the jet surface brightness diminishes along its length. The 1.4-GHz 
core power is $3.3 \times 10^{22}$~W~Hz$^{-1}$ and is only 0.36\% of 
the total power. In comparison with extended radio sources 
\citep{gi88}, the core power in MSH~05$-$2{\it 2} is a factor of four 
smaller than that typical for sources with the same total power; 
however, this deficit is small and within the observed dispersion in 
the relationship. A low core fraction is consistent with the 
edge-brightened structure, and an FR~{\sc ii} classification, but we 
note that the value of the core fraction estimated for 
MSH~05$-$2{\it 2} is 2--3 times below the median core fraction 
observed in powerful giant radio galaxies in \citet{Sch00} and 
\citet{Sa05}.

We show, in Fig.~10, slice profiles made across the two lobes of 
MSH~05$-$2{\it 2}. The slices have been made avoiding the peaks in 
total intensity. As a comparison, we show in Fig.~11 similar slice 
profiles made across PKS~J0116$-$4722 \citep{Sa02}, where we have 
similarly avoided the peaks in the brightness distribution over the 
lobes. Radio galaxy lobes that are actively supplied with energetic 
electrons via termination shocks in hotspots at the ends of jets are 
expected to be overpressured with respect to the ambient medium and 
the lobe expansion is limited by the ram pressure against the dense 
thermal environment.  Consistent with this picture, most extended 
powerful radio galaxies that we know of are observed to have well 
bounded lobes. Even in the case of powerful radio sources in which 
the jets have been interrupted for a brief period---as is the case 
in PKS~J0116$-$4722---we observe that the lobes are docked, as seen 
in Fig.~11. In contrast, the slice profiles across the lobes of 
MSH~05$-$2{\it 2} appear relaxed and the surface brightness 
diminishes gradually towards the edges; as far as we know this is 
the first such example in which the lobes of an edge-brightened 
radio galaxy are not sharply bounded. It may be noted here that this 
aspect of the source brightness distribution caused special 
difficulties during the image deconvolution, since the source 
extent---and, therefore, the `clean region' constraint---could not 
be easily specified {\it a priori}. The filamentary structure in the 
lobes of MSH~05$-$2{\it 2}, as seen in Fig.~2 and in the grey scale 
image in Fig.~10, and the significant intensity variations seen in 
the slice profile in Fig.~10, are also indications of significant 
entrainment that might be expected when relict lobes relax.

\subsection{The spectrum of radio lobe emission}

In Table~1 we present a compilation of flux density measurements of 
MSH~05$-$2{\it 2}. The 34.5~MHz measurement from the GEETEE telescope 
survey represents the integrated flux density and was derived from 
the survey images by first subtracting the Galactic foreground and 
then accounting for the fact that the source is resolved by the 
$57\farcm5 \times 26\arcmin$ beam. The value at 408~MHz was 
similarly derived from the 408-MHz survey image; in this case the 
quoted error is dominated by the uncertainty in the subtraction of 
the Galaxy.  The Parkes-MIT-NRAO (PMN) multi-beam survey images were 
used to estimate the flux densities at 4850~MHz. The radio spectrum 
of the source is a remarkably straight power law with index $\alpha 
= -1.05$ between 85.5 and 10700~MHz.  Between 34.5 and 85.5~MHz the 
spectrum appears flatter with $\alpha = -0.6$ at these long 
wavelengths.

In Fig.~12, we show the distribution in spectral index over the two 
lobes; the 1520-MHz image was smoothed to the resolution of the 
843-MHz image to make this comparison. Since the 1520-MHz image was 
constructed from full-synthesis data in the VLA CnB array that well 
images structures up to $15\arcmin$ in angular size, the smoothing 
to the MOST image resolution of about $1\arcmin$ is expected to 
yield reasonably accurate images for the comparison. Over a large 
region of the northern lobe (including the two peaks) the spectral 
index is steep and fairly uniform with $\alpha = -1.1$. Towards the 
dip in total intensity close to the center of the northern lobe the 
spectral index is distinctly steeper with $\alpha = -1.3$. All along 
the leading rim of the northern lobe the spectral index is 
significantly steeper, with $\alpha \approx -1.7$. In the southern 
lobe, the spectral index distribution is complex: $\alpha \approx 
-0.9$ towards the peak in the lobe and there is indication for 
relatively flatter spectral indices of $\alpha \approx -0.6$ towards 
the end of the southern lobe. Additionally, along the eastern and 
western edges of the southern lobe we observe patches with steep 
spectra of $\alpha \approx -1.5$.  Within the errors in the 
continuum imaging, we infer an unusual spectral index distribution 
asymmetry in this giant radio source. Profiles of the spectral index 
along the source axis show opposite behavior in the two lobes, with 
indices that gradually steepen to the north from the end of the 
southern lobe and gradually flatten to the south from the end of the 
northern lobe. While the observed trend in the southern lobe is 
typical, the northern lobe is extraordinary in that the spectral 
index flattens towards the core contrary to that observed in 
edge-brightened double radio sources. The spectral index is 
fairly constant over a $3\arcmin$ central region of both lobes.
 
\subsection{Synchrotron pressures in the radio lobes}

We first adopt here the minimum energy formalism \citep{Miley80}. We 
assume that the lobes of MSH~05$-$2{\it 2} have cylindrical symmetry 
and that the source axis makes an angle of $60\degr$ to the line of 
sight. The lobe plasma is assumed to be composed of electrons and 
positrons so that the ratio $(1+k)$ of the energy density due to all 
particles to that due to relativistic electrons is unity; 
evidence in favor of this assumption is given in \citet{Wardle98}. 
The lobe volume filling factor $\eta$ is assumed to be unity. 
Particle momenta are assumed to be randomly 
distributed in the lobes. We have made several slice profiles across 
both lobes, fitted Gaussians, and computed the peak surface 
brightness and lobe widths for the low surface brightness diffuse 
lobe emission. As discussed above, the slice profiles across the 
diffuse lobes do not exhibit sharp edges: the projected widths of 
the lobes are approximately twice the 
FWHM of the fitted Gaussians. In the formalism being considered 
here it is commonly assumed that the lower and upper cutoff 
frequencies for the emissivity are 0.01 and 100~GHz, respectively, 
in the rest frame and that the spectrum is described by a single 
power-law in between. In MSH~05$-$2{\it 2} the integrated spectral 
indices are observed to be $\alpha = -0.6$ below 0.1~GHz and $\alpha 
= -1.05$ at higher frequencies. In order to compute a lower limit to 
the pressure in the lobes within the minimum energy formalism, we 
assume a lower cutoff frequency of 0.1~GHz and power-law emission 
spectrum with $\alpha = -1.05$ above; the resulting magnetic field 
$B_{me} = 0.105$~nT. The corresponding lobe energy density is 
$u_{me} = 1.0 \times 10^{-14}$~J~m$^{-3}$.  Assuming that the magnetic field
lines are tangled, we estimate that the pressure in the 
diffuse parts of the lobes is $p_{me} = 3.4 \times 
10^{-15}$~N~m$^{-2}$.  It may be noted here that the radio polarization data
(Figs.~4 \& 5)
suggests that the fields are highly organized, particularly at the boundaries
of the lobes: as a result the effective magnetic pressure normal to the lobe
boundary might approach the magnetic energy density and the effective lobe
pressure might exceed the above estimate by a factor of about 1.5.
Since we have excluded the electrons that 
radiate in the 0.01--0.1~GHz range from consideration, our derived 
minimum pressures are an underestimate; however, we have estimated 
that the true pressure corresponding to the standard minimum energy 
assumptions, including electrons radiating down to 0.01~GHz, would 
exceed our derived value by less than a factor of 2. Additionally, 
deviations in the true values of $k$ and $\eta$ from those assumed 
above would increase the derived pressure; therefore, the value of 
pressure derived here is a true minimum.

We may instead adopt the equipartition formalism that assumes equal 
energy in particles and magnetic field; this formalism is perhaps 
appropriate for the diffuse regions of a relict lobe where 
relativistic particles and magnetic fields might be sufficiently 
coupled for energy exchange to result in equilibrium. The scaling 
relations in \citet{Leahy91} lead to estimates for $B_{eq}$ that are 
a small factor 1.09 greater than the $B_{me}$ given above; the 
equipartition energy densities $u_{eq}$ and pressures $p_{eq}$ would 
exceed the values derived above by a factor of 1.2. This equipartition
pressure, $p_{eq} = 4.1 \times 10^{-15}$~N~m$^{-2}$, is used in the following
sections as a lower limit to the pressures in the lobes.

For the magnetic field derived above, lower frequency cutoffs of 
0.01 and 0.1~GHz correspond to the exclusion of electrons with 
Lorentz factor $\gamma$ less than $1.5 \times 10^{3}$ and $5 \times 
10^{3}$ respectively. The electron population in the diffuse lobes 
might very well extend down to lower $\gamma$ and \citet{Brunetti97} 
give expressions for equipartition magnetic fields as a function of 
a minimum cutoff, $\gamma_{min}$. The low frequency emission 
spectrum has a spectral index $\alpha = -0.6$. If the low energy 
electron spectrum, which we estimate to have an energy index $\delta 
= 2.2$ (where $N(\gamma) \propto \gamma^{- \delta}$), were to extend 
down to $\gamma$ of order unity, the lobe pressures would be greater 
than the above minimum energy estimates by factor approximately 3.

\citet{Beck05} have derived revised expressions for the 
equipartition energies and magnetic fields in synchrotron plasma 
assuming that the relativistic particles are generated by an 
electromagnetic acceleration mechanism that produces a power-law 
particle momemtum spectrum from a reservoir of low energy particles.  
Adopting our assumption that all of the energy is in the light 
radiating particles, the formulae yield an estimate of about 0.17~nT 
for the field and the corresponding equipartition pressure is 
$p_{eq} = 7.7 \times 10^{-15}$~N~m$^{-2}$.

\subsection{A scenario for the radio source evolution}

Injection spectra in powerful radio sources are believed to have 
$\alpha = -0.5$ to $-0.7$ corresponding to the spectral indices of 
active hotspots.  The spectrum of MSH~05$-$2{\it 2} suggests a break 
around 0.1~GHz, with $\alpha$ of $-0.6$ below and steepening to 
$-1.05$ above. In a relict radio lobe, significant steepening with 
$\Delta \alpha$ of at least unity may be expected owing to 
synchrotron aging after reacceleration ceases. However, there is no 
evidence for any additional break or further steepening up to a 
frequency of 10~GHz.

Spectral aging in the relict lobes of MSH~05$-$2{\it 2} would be 
expected to result in breaks in the emission spectrum, which evolve 
with time towards lower frequencies; the break frequency is given by
 $$
 \nu_{T} = 1.12 \times 10^{3} { {B} \over {(B^{2}+B_{\rm MB}^{2})^{2}t^{2}} }~{\rm GHz},
  \eqno(1) 
 $$
 where $B = B_{eq} = 0.114$~nT is the magnetic field strength in the 
lobes, $B_{\rm MB} = 0.343$~nT is the magnetic field strength 
equivalent to the cosmic microwave background energy density, at the 
redshift of the source, and $t$ (in Myr) is the age of the 
synchrotron lobes corresponding to the break frequency $\nu_{T}$.  
The total spectrum of the source has a break at about 0.1~GHz, where 
the spectral index $\alpha$ steepens by about 0.45, and any further 
steepening occurs beyond 10~GHz. We infer that the source spectral 
age is 0.3~Gyr and that the evolution in the relict phase has been 
for less than 0.03~Gyr or 10\% of the source lifetime.  It may be 
noted here that spectral ages estimated for radio galaxies are 
believed to be lower limits \citep{Scheuer95,Blundell00}.

The phenomenology presented above suggests the following picture. 
MSH~05$-$2{\it 2} is a relict that was once a powerful radio source.  
In the active phase, the jets encountered an anisotropic medium, 
with greater gas densities to the north and, consequently, the jets 
would have advanced with lower speeds to the north, resulting in a 
lobe length asymmetry. After the jets ceased feeding the lobes, the 
hot spots would have dissipated and the lobes would have expanded and 
relaxed to equilibrium with the ambient gas pressure. This has 
happened in a relatively short time as compared to the lifetime of 
the source, which is to be expected given the rarity of such relict 
radio sources.  Significant entrainment of ambient thermal gas might 
be expected during this evolution.

This relict radio source is observed to have an unusually low 
fractional core power, and the twin jets close to the core appear to 
lack the collimation usually observed in edge-brightened radio 
sources.  We are led to believe that the central AGN is being 
observed following a relatively rapid transition to a low-power 
phase in which the jets rapidly expand, drop in surface brightness, 
and dissipate in the vicinity of the core without a termination 
shock.

The observed offset of the northern lobe from the axis defined by 
the core, the two jets observed close to the core, and the southern 
lobe, suggests that the relict northern lobe has been displaced to 
the west or SW, presumably after the jets stopped feeding the lobes. 
The observed asymmetries are unlikely to be due to a motion of the host galaxy
away from the centroid because (i) the source does not have a symmetry axis through
the centroid, and (ii) the low value for the galaxy overdensity in the
vicinity of the radio source indicates that the evolution of the large scale
structure surrounding the radio source is at the stage of developing
non-linearities, the gas is yet to be virialized, and we may expect 
that the gravitational motion of the galaxy and gas is the same.

The unusual steepening of the spectral index observed along the 
leading rim of the northern lobe---discussed in Section~5.1--- may 
be related to this 
displacement. There are arguments suggesting that the lobe material 
is composed of a multi-phase medium consisting of particles with a 
range of $\gamma$, and that the high-$\gamma$ particles might 
potentially be transported relative to a matrix defined by the 
lower-$\gamma$ particles \citep{Blundell00}.  In this case, we might 
expect that displacement of a lobe owing to forces arising from, for 
example, buoyancy that acts from without and within (as a result of 
instabilities and mixing) would result in a segregation in the 
electron energies: particles with high-$\gamma$ would suffer greater 
displacement relative to lower-$\gamma$ material that has higher 
inertia. Such a segregation could result in variation in electron 
energy distribution along the force axis, which would result in a 
spectral flattening along the direction of the external force.

\section{The physical state of the ambient medium of MSH~05$-$2{\it 2}}

As discussed above, we have reasons to believe that the lobes of the 
radio source MSH~05$-$2{\it 2} are relicts that have relaxed to 
equilibrium with the ambient IGM; therefore, we expect that the 
thermal gas pressure in the Mpc-scale environment of the radio 
source is at least equal to the equipartition pressure $p_{eq} = 4.1 \times 
10^{-15}$~N~m$^{-2}$ estimated earlier.

\subsection{Galaxy overdensity in the vicinity of MSH~05$-$2{\it 2}}

As discussed in Section~4, the host galaxy of the radio source 
MSH~05$-$2{\it 2} lies at the SW edge of a large-scale galaxy sheet 
that extends to lower redshifts. To infer the large-scale 
distribution in galaxy overdensity in the vicinity of the radio 
source, we have computed the 3D spatial distribution in the number 
density of the 6dF survey galaxies within a cube of side 60~Mpc 
centered at the host. Galaxies within $\pm 11\fdg11$ in R.A. and declination, 
and within $\pm0.0072$ in redshift space, were included in this 
analysis.  Using a spherical `top-hat' smoothing function of radius 
$R=5$~Mpc, the rms fluctuation in galaxy number density was found to 
be 1.7 times the mean.  The peak value of the fractional galaxy 
overdensity within the 60-Mpc cube was $\Delta n / \overline{n} = 
14.3$; this peak was located in the galaxy sheet to the NE of the 
host galaxy and at a position offset 5.4 Mpc east, 10.8 Mpc north 
and $-19.5$~Mpc along the line of sight, with respect to the host 
galaxy. At this smoothing scale, the fractional galaxy overdensity 
at the location of the host is $\Delta n / \overline{n} = 2.9$. On 
varying the smoothing scale $R$ in the range 3.7--10~Mpc (in which 
the mean galaxy number density always exceeds unity), the peak value 
of the fractional galaxy overdensity in the sheet varied in the 
range 18--5, and the fractional overdensity at the location of the 
host galaxy, which is at the edge of the sheet, varied in the range 
5.7--1.5.

Apart from this galaxy sheet that is located to the NE and in the 
foreground of the radio source, the 2dF data showed a relatively 
smaller galaxy concentration to the north of the radio source. We 
have merged the lists of 2dF (our observations presented in 
Section~4) and 6dF galaxies within $\pm 1\degr$ of the host galaxy, 
and used this database to determine the local fractional density 
variations in the immediate environment of the host.  Contour 
representations of this fractional overdensity distribution in
redshift slices of $\pm 0.00024$ are shown in Fig.~13; we have used a spherical 
`top hat' smoothing function with $R=1$~Mpc.

The host galaxy is itself part of a small group of galaxies that are 
distributed within about $6\arcmin$ (0.3~Mpc) on the sky; this group 
appears as a galaxy overdensity centered at the host in Fig.~13. Our 
redshift survey catalogs five members of this group: they have a 
mean velocity of $+20$~km~s$^{-1}$ with respect to the host galaxy 
and the standard deviation in the velocities is $83$~km~s$^{-1}$. 
Using an $R=1$~Mpc smoothing radius, this small group that has a low 
velocity dispersion represents a factor 3.1 overdensity.   
Comparing the properties of this group with samples in the 
literature \citep{Jo03,Kh07}, we infer that the low X-ray luminosity 
associated with the host, the optical luminosity, and the low 
velocity dispersion in the group classifies this group to be a low 
X-ray luminosity group and not a fossil group.

The nearest galaxy concentration outside this group is centered at a 
position with offsets 1.1~Mpc in RA, 1.35~Mpc in DEC, and 0.5~Mpc in 
redshift space; the corresponding linear separation from the 
position of the host galaxy is 1.8~Mpc ($40\arcmin$) towards NE, at 
a P.A. of $40\degr$.  The peak fractional overdensity has a value of 
$\Delta n / \overline{n} = 13$ for this system. In the same sky 
region, a second peak in overdensity is centered at a position with 
offsets 0.0~Mpc in RA, 0.8~Mpc in DEC, and 6.2~Mpc in redshift space 
(+450~km~s$^{-1}$). This overdensity has a peak $\Delta n / 
\overline{n} = 15$.  Both of these overdensities, which are located 
just north of the radio source, are connected and appear to be part 
of a filament running along the line of sight and extending in 
velocity space over the range $-170$ to $+750$~km~s$^{-1}$.  This 
structure has a mean velocity of $+250$~km~s$^{-1}$ with respect to 
the host galaxy and the standard deviation in the velocities is 
$230$~km~s$^{-1}$.

\subsection{The IGM environment of MSH~05$-$2{\it 2}}

Primordial nucleosynthesis \citep{Burles98} and CMB anisotropy 
\citep{Spergel06} suggest similar values for the baryon density in 
the Universe: $\Omega_{b} = 0.039$ and $\Omega_{b} = 0.042$ 
respectively. Estimates of the total baryon content at high 
redshifts ($z\approx 3$), most of which is expected to reside in 
Lyman-$\alpha$ absorbers, are in agreement with these expectations; 
however, all of the detected baryons in the present day 
Universe---including gas in clusters and that estimated for 
groups---accounts for only about half of the baryons 
\citep{Fukugita98}, leading to the problem of `missing baryons'. 
High-resolution cosmological hydrodynamic simulations 
\citep{Ce99,Da01} indicate that 40--50\% of the baryon content of 
the Universe---which is roughly the missing fraction---resides as 
WHIM gas associated with unvirialized overdensities.  The 
environment of MSH~05$-$2{\it 2}, including the small group of which 
the host is a member as well as the galaxy concentrations to the 
North and NE and galaxy sheet to the NE are all galaxy overdensities 
with $\Delta n / \overline{n}$ in the range 3--15, all in the regime 
of 
unvirialized overdensities, if the galaxies trace the total mass. 
Therefore, the IGM gas content of all of these structures observed in 
the vicinity of the radio source may be within the category of WHIM 
gas.

If we assume that $50\%$ of the baryon density of the Universe is in 
the IGM outside rich clusters, the mean IGM gas 
density is $\overline{\rho} = 2.2 \times 
10^{-28}$~kg~m$^{-3}$, corresponding to a mean particle density 
$\overline{n} = 0.26$~m$^{-3}$.  If we also make the assumption that 
the 
galaxies trace the gas density, the thermal pressure in the ambient 
gas in the vicinity of MSH~05$-$2{\it 2} is given by: 
 $$ p = 4.2 \times 10^{-17} \left( {{\Delta n}\over{\overline{n}}} + 
1 
 \right) \left( {k_{\rm B}T}\over{1~{\rm keV}} \right)~{\rm N}~{\rm m}^{-2}, 
 \eqno(2)
 $$ 
where $k_{\rm B}$ is the Boltzmann's constant.
Since we are considering the scenario where the lobes, with $p_{me} = 4.1 
\times 10^{-15}$~N~m$^{-2}$, are in equilibrium with the ambient gas, 
 we have the result that: 
 $$ \left( {{\Delta n}\over{\overline{n}}} + 1 \right) \left( 
 {k_{\rm B}T}\over{1~{\rm keV}} \right) > 100. \eqno(3)
 $$ 
 It may be noted here that if the radio lobe is not in 
equipartition, or if the filling factor departs from unity, or if 
there is significant energy in heavy particles, the IGM gas 
environment of the radio source would be constrained to have a 
higher density-temperature product. As discussed in Section~5.2, 
low-$\gamma$ electrons, if present, would additionally raise the value 
of this product. \citet{Dunn05} estimate $k/\eta$ in active and 
`ghost' lobes: $k/\eta$ may exceed unity by 3 orders of 
magnitude indicating that the true density-temperature product in 
the gas surrounding MSH~05$-$2{\it 2} may be as much as a factor 50
higher than that indicated by the above constraint.

The low velocity dispersions in the group, of which the host galaxy 
is a member, and that in the galaxy concentrations and galaxy sheet 
in the vicinity of the radio source, as well as the low fractional 
overdensity in these large-scale structures suggest that they are 
unvirialized, but dynamically evolving in the non-linear regime.  
The gas in these structures is expected to be primarily heated by 
hydrodynamic shocks and the temperature would correspond to the 
scales currently going non-linear.  Using the fits provided in 
\citet{Ce99}, the gas temperature is expected to be $k_{\rm B}T = 0.006 
(\lambda_{NL})^{2}$~keV, where $\lambda_{NL}$ (in Mpc units) is the spatial 
wavelength corresponding to the structure. The highest temperature 
shocks in the environment of the radio source would be those 
associated with the galaxy sheet, which has a scale length of about 
$8$~Mpc (corresponding to an angular size of $3\degr$).  This scale 
length roughly corresponds to the non-linear length scale for 
structure formation at $z=0$ \citep{valageas03}, and it is plausible 
that the shock-heated IGM gas temperature associated with this 
sheet could be in the vicinity of 0.4~keV. The hydrodynamic 
simulations of this heating indicate a scaling relationship for the 
WHIM gas temperature with fractional overdensity \citep{Da01,Ce06}: 
regions with overdensities below $10^{2}$ are not expected to be 
heated above about 0.4~keV. It may be noted here that 
ROSAT pointed observations of the Sculptor 
supercluster \citep{Zappa05} suggest IGM gas temperatures less than 
0.5 keV within the supercluster filaments.  
However, this heating, which is due to shocks close to the 
centre of the filaments as matter from the ends of the perturbation 
cross, would affect gas in only a narrow part of the filament a few 
hundred kiloparsecs wide; the peripheral regions of filaments, where 
MSH~05$-$2{\it 2} is located, may not be heated by this mechanism at 
$z=0$.

The lobes of the giant radio source are likely sampling the 
outer regions of the galaxy filament that are dynamically young.  
If the only source of heating has been the reionization of the IGM 
gas, then the ambient gas temperature might be as low as 1--10~eV, 
corresponding to a photoionized IGM that has been cooling with the 
expanding Universe.

Assuming hydrostatic equilibrium, the gas temperature associated 
with a galaxy clump with velocity dispersion $\sigma_{g}$ is of 
order $0.064 (\sigma_{g}/100~{\rm km}~{\rm s}^{-1})$~keV.  The 
observed velocity dispersions indicate that any virialized gas in 
the potential wells of the group as well as the nearby galaxy 
concentrations to the NE and north would have temperatures well 
below 0.5~keV. 

We are led to the conclusion that the gas in the vicinity of the 
radio source, which is observed to have an overdensity factor less 
than about 10 on the scale of the radio source (projected linear 
size of 1.8~Mpc) has a temperature less than 0.5~keV.  This results 
in a density-temperature product (or pressure) that is an order of 
magnitude below that expected from the properties of the radio 
source.

\section{Dynamics in the radio lobes of MSH~05$-$2{\it 2}}

The double lobe structure of MSH~05$-$2{\it 2} is asymmetric: the 
extents of the lobes on the two sides differ and the lobes are not 
collinear with the central core and jets. If the jet powers were the 
same on the two sides and if the interstellar medium of the host 
galaxy was not the cause for asymmetry in the jets emerging into the 
IGM, then the large-scale asymmetries in the radio galaxy may be 
related to gradients and anisotropies in the ambient IGM. The 
asymmetric structure is indicative of an evolution in which the jets 
advanced with different speeds on the two sides, resulting in 
different lobe extents, and after the jets stopped feeding the lobes 
there has been a movement of at least the northern lobe away from the 
source axis and towards SW. The observed extension past the core in 
the backflow associated with the northern lobe is consistent with 
such a picture.

We now discuss plausible mechanisms for the generation of the observed
asymmetries. 

\subsection{Asymmetries in gas density}

The observed ratio of the extents on the two sides suggests that the 
hotspot advance speeds on the two sides were in the ratio 1.6:1 or 
smaller.  Since the hotspot advance is ram pressure limited by the 
ambient gas density, the density ratio on the two sides might be as 
much as 2.25:1 on a scale of 1--2~Mpc.  As seen in Fig.~13, 
Panel~(b), the fractional overdensity to the north of the source is 
significantly greater than that to the south.  Smoothed to a scale 
$R=1$~Mpc, the galaxy density---and therefore perhaps also the gas 
density---at a distance 0.8~Mpc to the north is a factor 3--4 higher 
as compared to that to the south of the host galaxy.  The gas 
density gradient indicated by the galaxy density gradient appears 
consistent with that required to constrain the growth of the 
radio source to the north in comparison to the south.

{\subsection{Buoyancy}

We have used the 2dF and 6dF galaxy distribution data to infer the 
gravitational acceleration vector at the location of the radio 
source, due to the large scale distribution of matter.

The 6dF galaxies in a cube 60~Mpc a side---$22\fdg22$ wide in R.A. and 
declination and $\pm 0.0072$ in redshift 
space---were included in the first analysis.  This amounted to $N_{g} = 
1004$ galaxies.  The total matter density in the cube was assumed to 
be the volume $V$ of the cube times the mean matter density in the 
Universe: $M_{tot} = V \times \rho_{m} = V \times \Omega_{m} 
\rho_{c}$, where the critical density $\rho_{c} = 0.94 \times 
10^{-26}$~kg~m$^{-3}$. Assuming that the galaxies trace this matter, 
every one of these 6dF galaxies was assigned a mass $M_{g} = 
M_{tot}/N_{g} = 7.7 \times 10^{12}$~M$_{\odot}$. Neglecting the 
matter outside the cube, the acceleration vector owing to the 1004 
galaxies within the cube was computed; the components are: $g_{\rm 
RA} = 1.24 \times 10^{-12}$~m~s$^{-2}$, $g_{\rm DEC} = 7.66 \times 
10^{-13}$~m~s$^{-2}$ and $g_{z} = 3.88 \times 10^{-13}$~m~s$^{-2}$ 
along increasing R.A., declination and redshift respectively.  The magnitude 
of this acceleration vector is $\vert g \vert = 1.5 \times 
10^{-12}$~m~s$^{-2}$, and assuming the age of the Universe to be 
$\tau = 13.7$~Gyr, we infer that the environment of the radio source 
might be experiencing an infall towards the sheet to the NE with a 
speed about $\vert g \vert \times \tau = 650$~km~s$^{-1}$.  The 
acceleration vector also implies that the environment of the radio 
source might have a line-of-sight peculiar velocity of about 
170~km~s$^{-1}$; if so, the true location of the host galaxy and the 
group it is associated with might be 17.1~Mpc---instead of 19.1~Mpc 
(see Section~6.1)---along the line of sight from the peak 
overdensity in the galaxy sheet, which is located in the foreground 
and to the NE of the radio source.  On the sky plane, the 
acceleration $g_{s}$ is directed towards P.A. of $60\degr$, with 
a magnitude corresponding to $g_{s} \tau = 630$~km~s$^{-1}$.

The local acceleration vector was determined using the 2dF data.  A 
total of 21 galaxies in a 5.4~Mpc wide cube---$2\degr$ wide in R.A.
and declination and $\pm 0.000648$ in redshift space---were included in this 
analysis. Following the same assumptions and procedure described 
above, each galaxy was assigned a mass of $2.7 \times 
10^{11}$~M$_{\odot}$. The resulting gravity vector was computed to 
be: $g_{\rm RA} = 2.53 \times 10^{-14}$~m~s$^{-2}$, $g_{\rm DEC} = 
9.60 \times 10^{-14}$~m~s$^{-2}$ and $g_{z} = 3.50 \times 
10^{-14}$~m~s$^{-2}$.  On the sky plane, this corresponds to $g_{s} 
\tau = 43$~km~s$^{-1}$ towards P.A. of $15\degr$. In order to 
compute the acceleration experienced by the giant radio source, 
which has a 1.8-Mpc projected linear size, we have neglected 
galaxies within 1~Mpc of the host galaxy when computing these 
acceleration vectors. The corresponding redshift-space distortion 
amounts to a line-of-sight peculiar velocity of 15~km~s$^{-1}$, or 
errors of 0.2~Mpc in linear distance, which is small compared to 
the distances to nearby galaxy concentrations.

The gravity gradient vector direction computed above suggests that a 
possible mechanism for the movement of the northern radio lobe 
towards SW or south is buoyancy: in this model, the light 
synchrotron lobes would rise against the gravity vector direction as 
a result of being enveloped by thermal IGM gas that is significantly 
denser.  This movement might effectively take place during the time 
since the jets ceased feeding the lobes.  The model being considered 
here is similar to that in clusters of galaxies, where X-ray `holes' 
in the intracluster gas are believed to be synchrotron plasma 
`bubbles' rising buoyantly in the dense thermal gas associated with 
the cluster \citep{Chu00}. The terminal speed of the movement of the 
northern lobe, as a result of buoyancy, would be given by:
 $$
 v_{\rm T} = \sqrt{ {2gV}\over{sC_{\rm D}} }, \eqno(4)
 $$
 where $g$ is the acceleration due to gravity, $V$ is the volume of 
the northern lobe, $s$ is the cross section area, and $C_{\rm D}$ is 
the drag coefficient, which we assume to be 0.75 \citep{Chu01}.  
For the northern lobe, $(V/s) \approx 360$~kpc. Even using the 
relatively large acceleration, corresponding to $g_{s} \tau = 
630$~km~s$^{-1}$, which was estimated for the large scale structure, 
the terminal velocity is $v_{\rm T} = 208$~km~s$^{-1}$. A 
displacement over a distance of 180~kpc, corresponding to half the 
width of the northern lobe, would require about 1~Gyr.  The 
acceleration corresponding to the relatively nearby galaxy 
concentrations to the north were computed above to be a factor 15 
smaller, and the terminal velocity due to this gravity vector would 
be a factor of 3.8 slower and the rise time correspondingly longer.

The $t = 1$~Gyr timescale we have computed above for the buoyant 
displacement of the northern lobe from the source axis is somewhat 
longer than the spectral age of the source computed in Section~5.3, 
which may itself be an underestimate of the true dynamical age of 
the radio lobes.  The conclusion is that buoyancy forces in the IGM 
are a plausible mechanism for the observed displacement provided that the
displacement occurs over the total source lifetime; however, 
additional causes might also play a significant role.

\section{Discussion}

We derive a lower limit of $4.1 \times 10^{-15}$~N~m$^{-2}$ to the 
synchrotron pressure within the radio lobes; this suggests a 
density-temperature product of $3 \times 10^{8}$~K~m$^{-3}$ for the 
ambient IGM. The galaxy group, nearby galaxy concentrations, and the 
large scale sheet at the edge of which the radio source is located, 
all have fractional density contrasts of order 10 or smaller.  The 
radio source, which has a projected linear size of 1.8~Mpc, is
presumably interacting with the IGM associated with a filamentary large-scale
structure in the galaxy distribution: the 
derived density-temperature product for this structure 
is a measurement of the physical 
state of the IGM gas in the vicinity of the radio source.  
Assuming that galaxies trace the unseen IGM gas, the densities and 
temperatures we expect for the IGM environment---based on the local 
galaxy overdensity and assuming that the IGM gas is heated by 
hydrodynamic shocks associated with the formation of the observed 
large-scale structure---falls at least an order of magnitude short 
of the expectations based on the radio properties of the source. 

There are several plausible explanations for this basic result that the IGM
pressure is at least an order of magnitude below that in the radio lobes, we
discuss some of them below.

\subsection{Do the galaxies in the filaments trace the gas?}

The distribution in the unseen collisionless dark matter dominates the
gravitational potential of the large scale structure.  Galaxies display
luminosity segregation \citep{Pe02}---the distribution of luminous galaxies
are biased---and, therefore, derivations of mass overdensity factors based on the
space distribution of luminous galaxies would be
overestimates.  Consequently, the gravitational potential and 
gravitational accelerations would also be overestimates. 

We have assumed that the galaxies trace the gas, and that the galaxy 
overdensity factors derived represent the IGM gas overdensity.  
Unlike dark matter and galaxies, gas is dissipative and it is possible that
the gas distribution is significantly more biased compared to galaxies.
However, this might only be expected in later stages of non-linear structure growth
when the overdensity factors attain values exceeding about $10^{2}$ and
radiative cooling becomes significant. In the
mildly overdense stages when the overdensity factor is a few 10's and the
structures are yet to collapse, we might expect the gas and galaxies to have
similar overdensity factors.  

\subsection{The ambient medium of the radio lobes: is it an intra-group
  medium?} 

The small group of which the host is a member has a 
relatively small velocity dispersion compared to that typically 
observed in groups of galaxies (see, for example, the compilation in 
\citet{Mu03}) and, therefore, it is expected that any virialized gas 
associated with the group would have a temperature at the lower end 
of the observed distribution: 0.4--1.5~keV. Groups with $k_{\rm B}T < 1$~keV 
are detected to have a gas extent less than half the virial radius of 
the galaxies in the group, or out to a radial distance of less than 
0.4~Mpc. It may be noted here that the lobes of MSH~05$-$2{\it 2} 
extend well beyond the obvious extent of the group, of which the 
host galaxy is a member.

Although it is unlikely that the gas associated with the small 
group---of which the host galaxy is the dominant member---extends 
over the entire radio source, this expectation needs to be 
confirmed with sensitive X-ray imaging of the environment of the radio
source. Based on the observations to date, it is reasonable to assume that the
radio lobes interact with the more diffuse IGM associated with the large-scale
filament and not thermal intra-group gas.

\subsection{Are the radio lobes in equilibrium with the IGM?}

Active radio sources inflate cocoons that are likely
overpressured and expanding with speeds limited by ram pressure balance.
In this model, 
the synchrotron cocoons are enveloped in gas that has been shock
heated and compressed by the mechanical 
action of the expanding radio lobe in the vicinity of the contact
discontinuity.  
Once the central beams are switched off, the relict radio sources are believed
to rapidly diminish in surface brightness---as evidenced by the paucity of
relicts---as the cocoons and enveloping thermal gas pressure equilibrates with
the IGM. 

An assumption made here is that the radio lobes of MSH~05$-$2{\it 2} are 
relicts and, therefore, in pressure equilibrium with the ambient thermal gas
outside the contact discontinuity.  In the light of the finding that the
pressure in the external IGM might be an order of magnitude underpressured, a
possible explanation is that the lobes of MSH~05$-$2{\it 2} are not
relaxed---despite the distinctive aspects of the radio properties that suggest
a relaxed relict state for the lobes---but are being observed in a rare
transition phase as a fading relict continuing ram-pressure
balanced expansion.  An observational support for this
picture is the enhanced and circumferential magnetic field along the boundaries of
the cocoons, which are indicative of radial compression for the synchrotron
plasma. 

In this interpretation, the lobes of MSH~05$-$2{\it 2}, with internal pressure
exceeding $4.1 \times 10^{-15}$~N~m$^{-2}$, are over pressured compared to the
external thermal gas, with density less than $2.2 \times
10^{-27}$~kg~m$^{-3}$ and pressure about $2.3 \times 10^{-16}$~N~m$^{-2}$. 
Ram pressure expansion with speed $v/c \approx 0.0045$ over a timescale
0.3~Gyr, which is similar to the lifetime of the source, would expand the
lobes by factor two and result in a relaxed relict in static pressure
equilibrium with the IGM associated with the large-scale filament in which the
source is embedded.  The surface brightness of the relaxed relict at 1.4~GHz would be
less than 0.25~mJy~arcmin$^{2}$, much less than the current surface brightness
of about 30~mJy~arcmin$^{2}$ and well below surface brightness detection
limits of wide-field surveys. This scenario for the evolution in
MSH~05$-$2{\it 2} would be consistent with the rarity of relaxed relict radio
galaxies---the `dead' radio galaxies (\citet{gi88}; see also the discussion in
\citet{Blundell00}). 

\subsection{Additional sources of heating of the IGM environment}

We discuss here the possibility 
that there is likely a different cause for 
the heating of the IGM gas in the vicinity of the radio source, 
which dominates over hydrodynamic shock heating or   
mechanical heating of the ambient gas by the radio source itself and might
permit the radio lobes to be in pressure equilibrium with the IGM.

The galaxy distribution in the vicinity of the radio source (Section~6.1)
shows a concentration of galaxies centered 
1.8~Mpc NE of the host galaxy that is significantly closer to the 
northern lobe than the southern lobe. 
The observational evidence for an association between the asymmetries in the
radio source and the large scale distribution of galaxies indicates that
a model worth considering is one in which shocks associated with 
galactic super winds (GSWs) or mechanical heating associated with any
AGNs in this galaxy concentration might enhance the
density-temperature product in the IGM environment of the radio source and,
additionally, also contribute to the
dynamical evolution of MSH~05$-$2{\it 2} and the displacement of the northern
lobe off axis and in the SW direction. 
  
Hydrodynamic simulations of the evolution in large-scale structure, 
incorporating 
such GSWs \citep{Ce06} have indicated the power of such winds in 
translating gas from higher density regions into lower density 
peripheral regions of filaments as well as driving shocks into 
regions with low fractional overdensities that would otherwise not 
have experienced shock heating during non-linear structure 
formation.  Apart from heating the ambient gas around the radio 
lobes of MSH~05$-$2{\it 2}, the GSWs might also transport gas from 
the galaxies in the concentrations and into the environment of the 
radio source.  In this picture, the energy arising from 
GSW or AGN related activity in the galaxies to the NE and  
propagating in the low density WHIM might 
aid the movement of the northern radio lobe towards the SW, as 
discussed in Section~7.

The estimated age 
of the relict source, $t = 0.3$~Gyr, implies a fairly large 
displacement speed of order 600~km~s$^{-1}$. \citet{Ru05} measure a 
median `maximum' velocity for the winds in their sample of starburst 
galaxies to be 350~km~s$^{-1}$; however, one galaxy in their sample 
was measured to have a wind speed of 1100~km~s$^{-1}$. \citet{He02} 
suggests that primary wind fluid velocities might be a few thousand
km~s$^{-1}$.  Additionally, the superwinds in the environment 
of MSH~05$-$2{\it 2} might be a combined wind from several of the 
galaxies in the nearby concentration to the NE. GSW speeds are, 
therefore, in the vicinity of the displacement speed inferred for 
the lobe. It may also be noted that in simulations, GSWs tend to 
move towards the low density regions (\citet{Ka07} and references 
therein). Since MSH~05$-$2{\it 2} is located between the galaxy 
concentration to the NE and a void to the SW, winds from the observed galaxy
concentration would be expected 
to displace the northern lobe in the required direction. Therefore, 
it is plausible that such astrophysical feedback mechanisms, 
which may be responsible for the high pressure 
in the ambient IGM, may also contribute to the radio source 
dynamics and result in the observed displacement of the northern lobe over the
source lifetime. 

We have looked for evidence of current AGN activity in the galaxies 
located in the nearby concentration to the NE of MSH~05$-$2{\it 2}, which are
within $\pm0.003$ of the host redshift.  A galaxy is coincident with a ROSAT
X-ray source 1WGA~J0506.8$-$2810, another may be identified with a radio
source in the  
NRAO VLA sky survey (NVSS) as well as a ROSAT object
1WGA~J0506.6$-$2814. About 10\% of the 
galaxies appear to be AGNs, which may be responsible for the feedback
mechanisms being considered here.

We have examined an archival ROSAT PSPC image of the sky region in the
vicinity of the radio source for evidence for extended emission on the scale
of the radio source.  This was a 22~ks pointed observation. 
First, smoothing the ROSAT image does not indicate
any evidence for extended X-ray emission associated with the group of which
the host galaxy is a member.
Second, if the double radio lobes were enveloped by 
relaxed hot thermal gas on a scale
similar to the size of the radio lobes, and with a pressure as indicated by
the radio lobes, the expected counts in the
0.2--2.0~keV ROSAT band would be below the sensitivity to such
structure. Existing data do not rule out the GSW or AGN feedback 
models considered in this section. 

Adopting an electron density of $n_{e} = 0.26$~m$^{-3}$ and a temperature of
0.5~keV (see Section~6.2), and the expression for temperature relaxation
timescale in RA: 05$^{h}$ 03$^{m}$ $51\fs0$, DEC: $-$28$\degr$ 
45$\arcmin$ 00$\arcsec$
\citet{Yo05}, we infer that the electron-ion equilibration timescale in the
ambient IGM gas is about 0.014~Gyr. A higher ambient density would reduce the
equilibration time; however, a higher ambient temperature resulting from
heating via shocks associated with GSWs or AGNs could potentially increase the
equilibration timescale significantly. It is, therefore, possible---although
unlikely---that in the 
case of a shock heated IGM the equilibration timescale exceeds the age
associated with the relict phase, as well as the age of the radio source. As a
consequence, we cannot rule out a model where the ambient IGM beyond the
contact discontinuity of the radio synchrotron plasma is composed of hot ions
and relatively cooler electrons, and hence difficult to detect in its
continuum radiation. 

\section{Summary}

We have made radio continuum images in total intensity and 
polarization of the $40\arcmin$ double radio source MSH~05$-$2{\it 2} 
with a beam of about $15\arcsec$ FWHM. Additionally, the extended 
low surface brightness emission has been imaged at 843~MHz and the 
radio core and associated jets at 4.8~GHz. The radio structure is 
highly asymmetric and there is evidence---in the morphology and 
other radio properties---that the lobes are relicts of a powerful 
radio galaxy.  We have examined the large-scale structure 
in the galaxy distribution around the radio source using 6dF survey 
data and our 2dF observations, with the aim of examining (1) whether 
gas associated with this structure might plausibly have the pressure 
indicated by the radio properties, and (2) whether the asymmetries 
in the radio source might be a consequence of the interaction 
between jets and lobes with gas associated with the galaxy 
distribution.  The radio source is at the SW periphery of a 100-Mpc 
scale galaxy sheet and we identify galaxy concentrations to the NE 
and North at distances 2--6~Mpc from the host galaxy.  The host 
galaxy is itself a member of a small group.

A relationship has been observed between the 
asymmetries in the radio morphology of
MSH~05$-$2{\it 2} and the large scale structure in galaxies. The observed 
asymmetries in the radio structure may be caused, in part, by the 
gas density gradient in the vicinity of the radio source, as 
inferred from the large scale galaxy density gradient. Additionally, 
buoyancy forces arising from the local gravity gradient due to the 
large scale matter distribution, and any winds from galaxy 
concentrations observed in the vicinity of the radio source, 
may together cause the displacement we infer for the 
northern lobe from the radio axis.

However, we find that the minimum pressure within the
radio lobes is at 
least an order of magnitude greater than the pressure expected for the gaseous
environment of the radio source.  This is surprising in the light of the
relationship between the asymmetries in the radio source and that in the
ambient large-scale structure, and the evidence that MSH~05$-$2{\it 2} is a
relict. We have discussed the possibility that the
relict radio lobes are not in pressure balance with the IGM associated with
the filamentary large-scale structure in galaxies. This explanation suggests
that extended radio sources located in low-galaxy-density environments may not
be relaxed relicts even if they have the properties of MSH~05$-$2{\it 2} that
indicate a relict state.  Relaxed relicts embedded in the large scale galaxy
filaments may have surface brightness---corresponding to the thermal pressures
in the IGM gas---that are below surface brightness limits of current surveys.
Surveys, like the Australia Telescope low-brightness survey (ATLBS;
\citet{Su08}), might detect the relaxed relict source population.

As a way of reconciling the relationship between the asymmetries in the radio
source, the evidence indicating that the lobes of MSH~05$-$2{\it 2} are relict,
and the mismatch between the internal pressure estimated for the synchrotron lobes and
the thermal pressure expected for the ambient IGM, we speculate that the
heating of the peripheral regions  
of the IGM associated with the galaxy filament surrounding the radio source 
might be dominated by 
astrophysical feedback from nearby galaxy concentrations, 
which heat gas located more 
than 2~Mpc away from galaxy concentrations. 
In this interpretation, MSH~05$-$2{\it 2} is a relaxed relict, and the
observations presented herein constitute  
independent evidence indicating that feedback processes from 
galaxies might dominate the state of the gas associated with
large-scale structure in the Universe.
It may be noted here that such 
extensive heating of the IGM via GSWs might also be the cause for 
the detection of extended Sunyaev-Zeldovich signals from clusters 
and groups \citep{Myers04}.

The conclusions we derive here are not without uncertainties.
Nevertheless, this work represents a first step towards using giant 
radio galaxies as a means of probing the IGM outside groups and 
cluster environments via a detailed comparison of radio source properties with
the 3D large-scale galaxy distribution.  

\acknowledgments

The MOST is operated by the University of Sydney and supported in 
part by grants from the Australian Research Council.  The National 
Radio Astronomy Observatory is a facility of the National Science 
Foundation operated under cooperative agreement by Associated 
Universities, Inc. Thanks are due to the Anglo-Australian 
Observatory for the data obtained using the 2dF instrument on the 
3.9-m Anglo-Australian Telescope, which was obtained in service 
time. We thank Will J. Sutherland for the RUNZ code.

\clearpage

\begin{deluxetable}{llrrrl}
\tablewidth{0pc}
\tablecolumns{6}
\tablecaption{Radio flux densities of MSH~05$-$2{\it 2}}
\tablehead{ \colhead{Freq. (MHz)} & \colhead{Survey/Telescope} & \colhead{N lobe  
      (Jy)} & \colhead{S lobe (Jy)} & \colhead{Total (Jy)} &
    \colhead{Reference} }
\startdata
34.5 & GEETEE & & & $102 \pm 5$ & \citet{dwaraka90} \\
85.5 & MSH & & & $60 \pm 12$ & \citet{mills60} \\
408 & 408-MHz all-sky survey & & & $11.8 \pm 1.2$ & \citet{haslam82} \\
843 & MOST & 2.82 & 2.04 & $4.9 \pm 0.25$ & This paper \\
1520 & VLA & 1.55 & 1.41 & $2.95 \pm 0.02$ & This paper \\
2695 & Effelsberg & 0.85 & 0.77 & $1.62 \pm 0.07$ & \citet{Sa88} \\
4850 & PMN & 0.45 & 0.38 & $0.83 \pm 0.04$ & \citet{tasker94} \\
10700 & Effelsberg & 0.193 & 0.17 & $0.369 \pm 0.011$ & \citet{jamrozy05} \\
\enddata
\end{deluxetable}

\clearpage

\begin{figure}
\epsscale{0.55}
\plotone{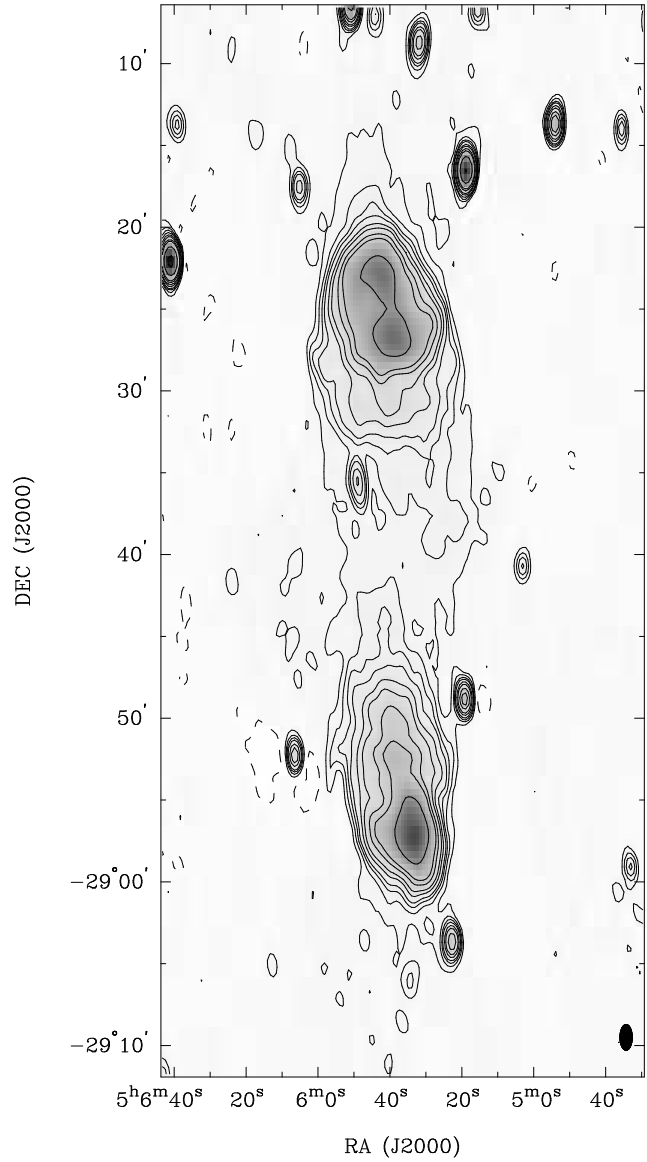}
\caption{ 
 843~MHz MOST image of MSH~05$-$2{\it 2}. The radio image has a beam 
of FWHM $94\farcs0 \times 45\farcs0$ at a P.A. of $0\degr$ and 
represents the co-addition of several archival images. Contours 
are at ($-$1, 1, 2, 3, 4, 6, 8, 12, 16, 32 and $64) \times 2.5$~ 
mJy~beam$^{-1}$; grey scales are over the range $-5$ to 200 
mJy~beam$^{-1}$. The lowest contour is at a level of 3 times the rms 
noise in the image.  In this image, as well as all following images 
displayed here, the half-maximum size of the beams of the radio 
images are shown using a filled ellipse in the lower right of the 
figure.  Additionally, all the radio images have been corrected for 
the attenuation due to the primary beam.
 \label{fig1}
	}
\end{figure}

\clearpage

\begin{figure}
\epsscale{0.7}
\plotone{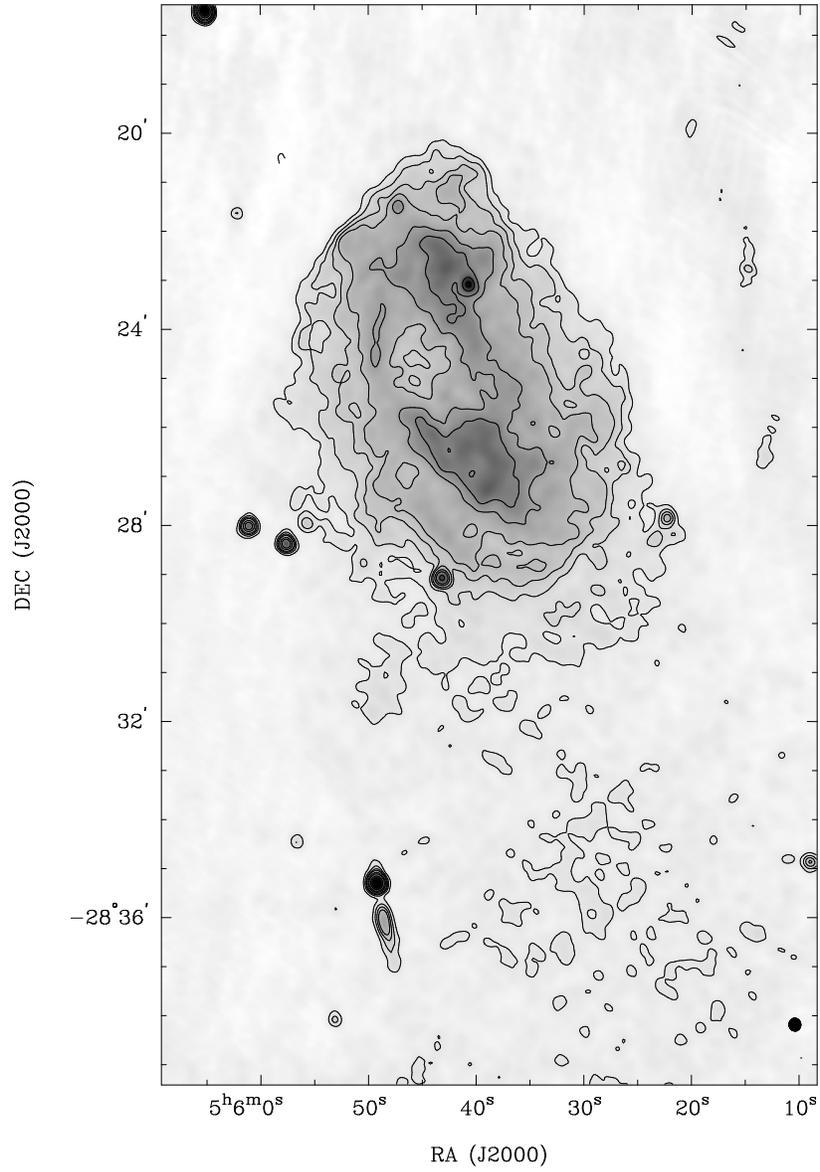}
\caption{ 1520~MHz total intensity VLA image of the northern lobe of MSH~05$-$2{\it 2}. 
The radio image has a beam of FWHM $15\farcs5 \times 14\farcs2$ at a P.A. of $8\fdg6$.
Contours are at ($-$1, 1, 2, 3, 4, 6, 8, 12 and $16) \times 
0.4$~ mJy~beam$^{-1}$; grey scales are over the range 
$-0.3$ to 7 mJy~beam$^{-1}$. 
\label{fig2}
	}
\end{figure}

\clearpage

\begin{figure}
\epsscale{0.7}
\plotone{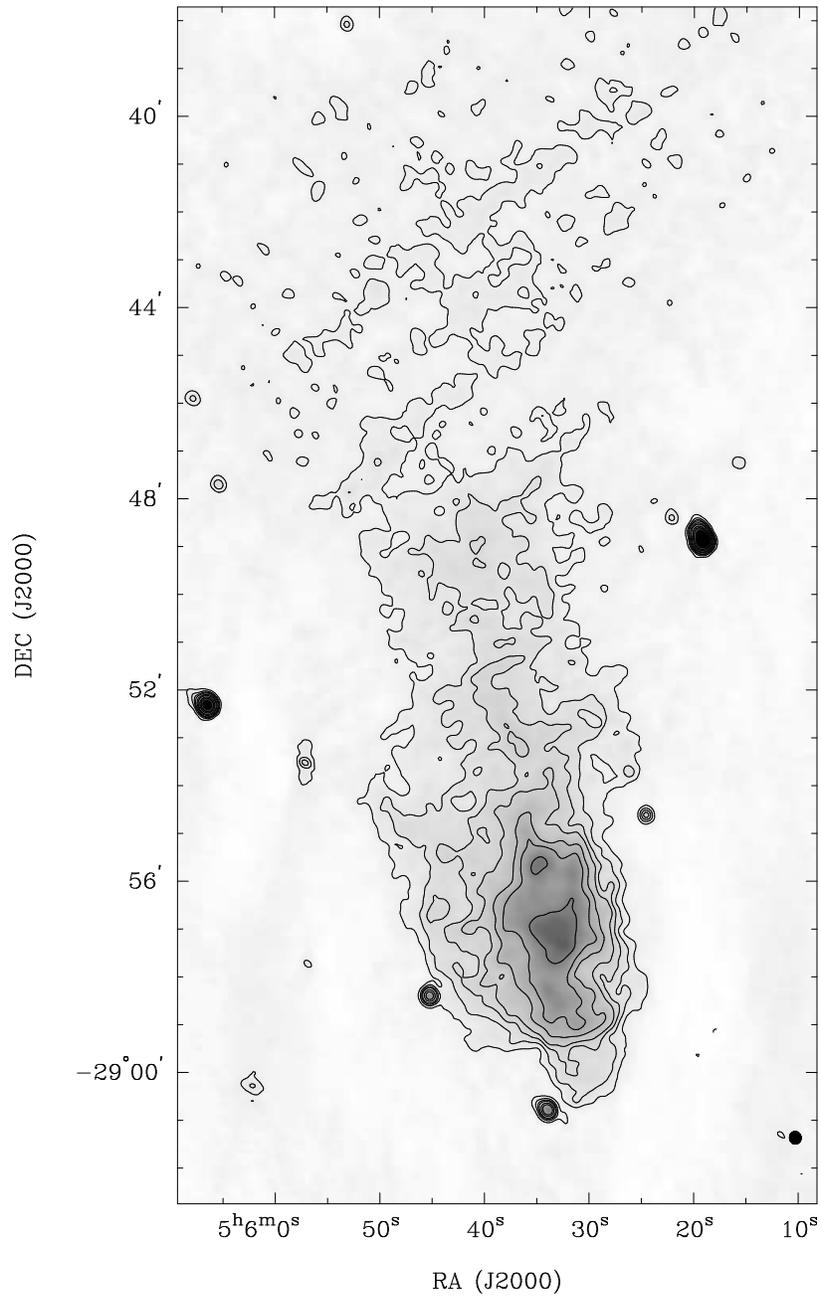}
\caption{ 
 1520~MHz total intensity VLA image of the southern lobe of 
MSH~05$-$2{\it 2}. The beam size, contour levels and grey scale range 
are the same as in Fig.~2.
 \label{fig3}
	}
\end{figure}

\clearpage

\begin{figure}
\epsscale{0.7}
\plotone{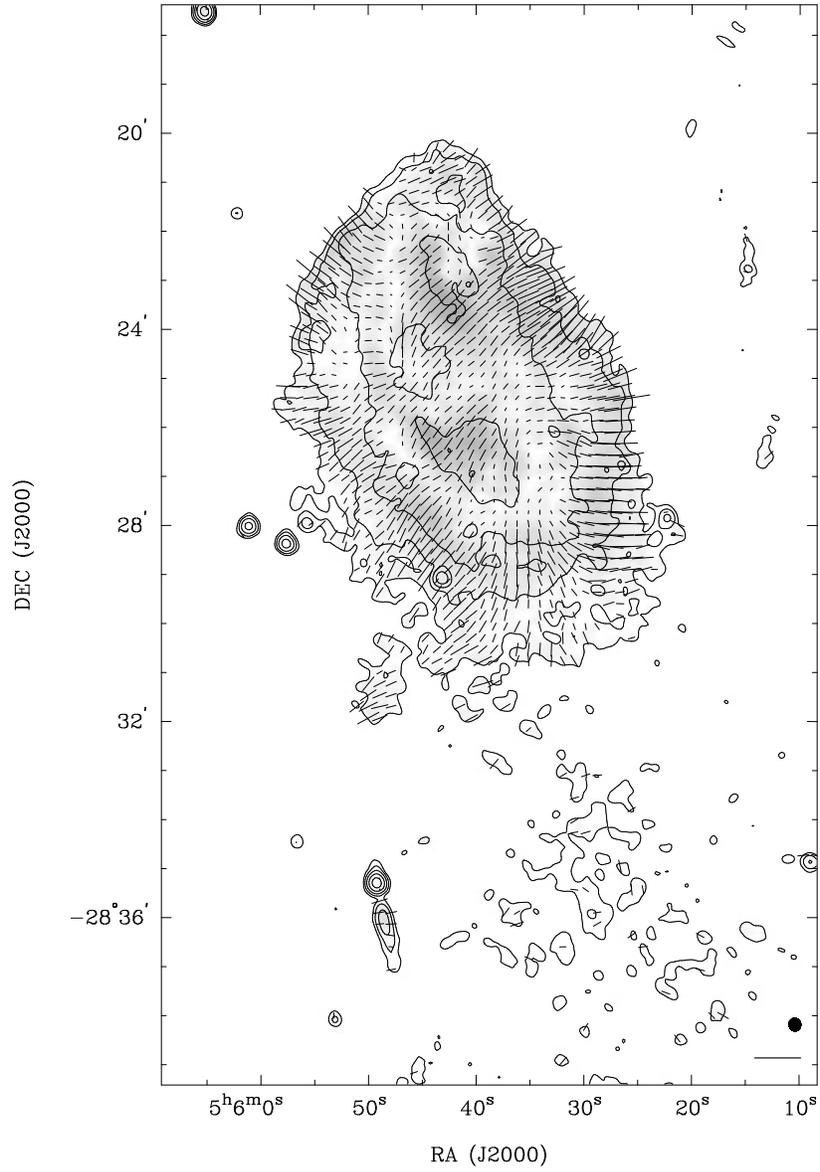}
\caption{ 1520~MHz VLA image of the polarization in the emission from the 
northern lobe of MSH~05$-$2{\it 2}. The polarized intensity is displayed using 
grey scales over the range 0--4 mJy~beam$^{-1}$. 
Contours of the total intensity are overlaid at ($-$1, 1, 2, 4, 8, 16, and $32) \times 
0.4$~ mJy~beam$^{-1}$.   Polarization vectors represent
the electric field orientation with length proportional to the fractional
polarization.  The length of the horizontal bar at the lower right 
corner of the image represents 100\% fractional polarization.  The polarization
vectors have been corrected for Faraday rotation along the line of sight.
\label{fig4}
	}
\end{figure}

\clearpage

\begin{figure}
\epsscale{0.7}
\plotone{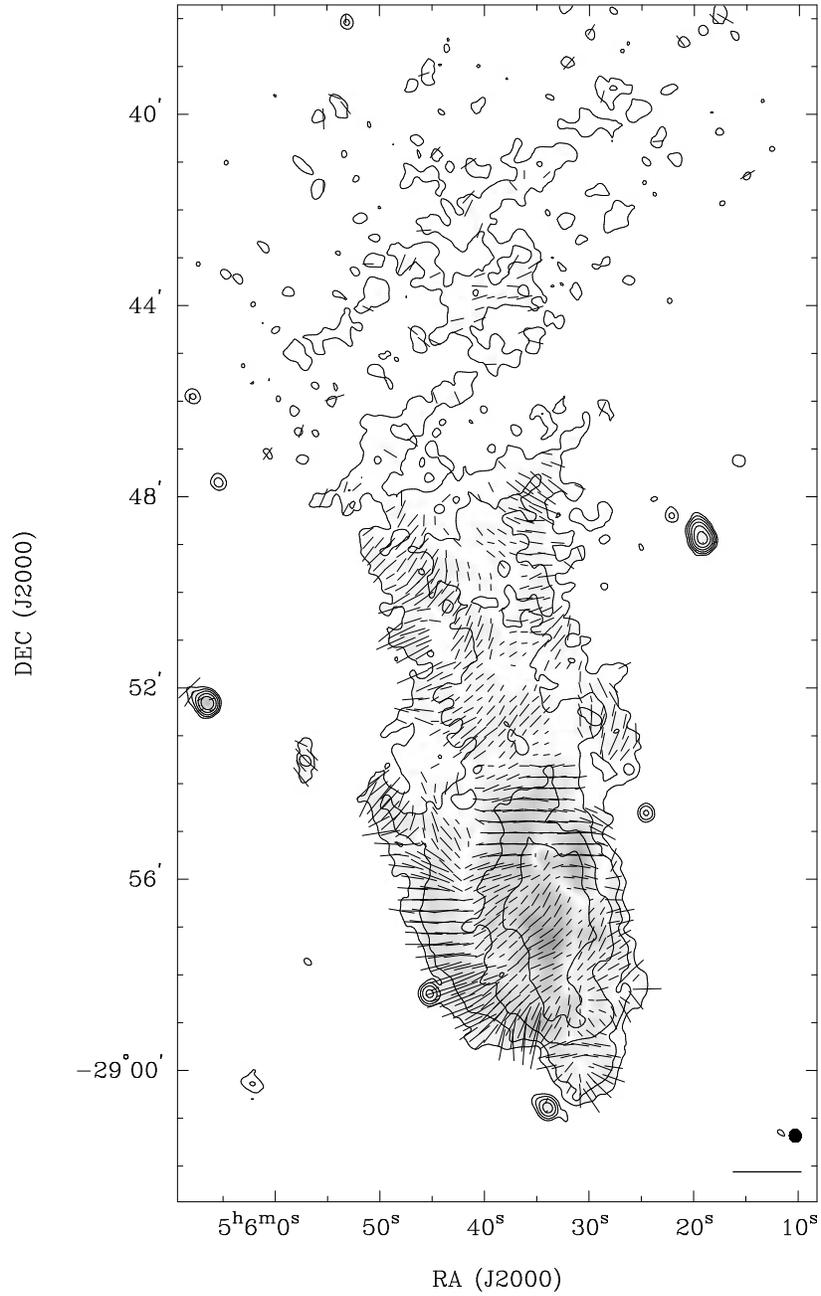}
\caption{ 
 1520~MHz VLA image of the polarization in the emission from the 
southern lobe of MSH~05$-$2{\it 2}.  The parameters for this image 
representation are the same as for Fig.~4.
 \label{fig5}
	}
\end{figure}

\clearpage

\begin{figure}
\epsscale{0.7}
\plotone{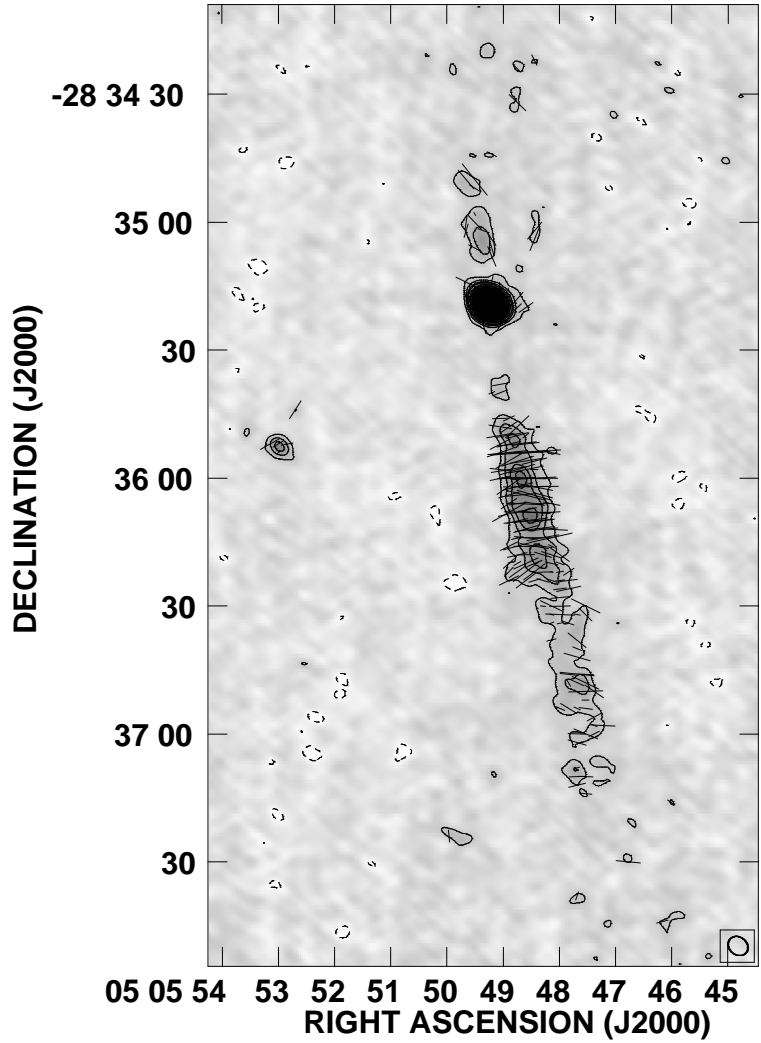}
\caption{ 
 4860~MHz VLA image of the radio core of MSH~05$-$2{\it 2}. The radio 
image has a beam of FWHM $5\farcs0 \times 4\farcs1$ at a P.A. of 
$50\fdg8$. The total intensity image is shown using contours and 
grey scales: contours are at ($-$1, 1, 2, 3, 4, 6, 8, 12, 16, and 
$32) \times 0.075$~ mJy~beam$^{-1}$ and grey scales are over the 
range $-0.1$ to 0.7 mJy~beam$^{-1}$.  The lowest contour is at 3 
times the rms noise. Polarization vectors represent the electric 
field orientation with length proportional to the polarized 
intensity: $1\arcsec$ length corresponds to 10~$\mu$Jy~beam$^{-1}$ polarized
intensity. 
 \label{fig6}
	}
\end{figure}

\clearpage

\begin{figure}
\epsscale{0.5}
\plotone{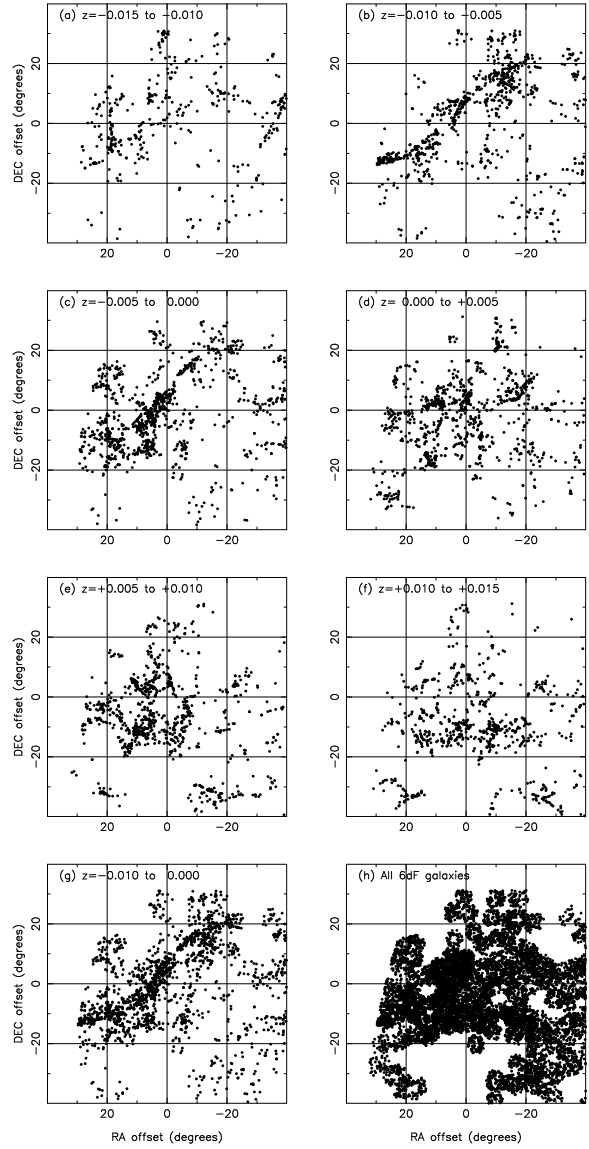}
\caption{ 
 The distribution of 6dF galaxies in the vicinity of 
MSH~05$-$2{\it 2}; $20\degr$ corresponds approximately to a linear 
distance of 54~Mpc. Panels (a) to (f) show the distribution in 
redshift slices that are 0.005 wide in redshift space and cover a 
redshift range $\pm 0.015$ about that of the host galaxy at 
$z=0.038286$.  Panel (g) shows all galaxies with redshifts in the 
range $-0.01$ and 0.0 relative to the host redshift.  Panel (h) 
shows all 6dF galaxies in this region with redshifts cataloged to 
be `reliable' or `probable'.
 \label{fig7}
}
\end{figure}

\clearpage 

\begin{figure}
\epsscale{1.0}
\plotone{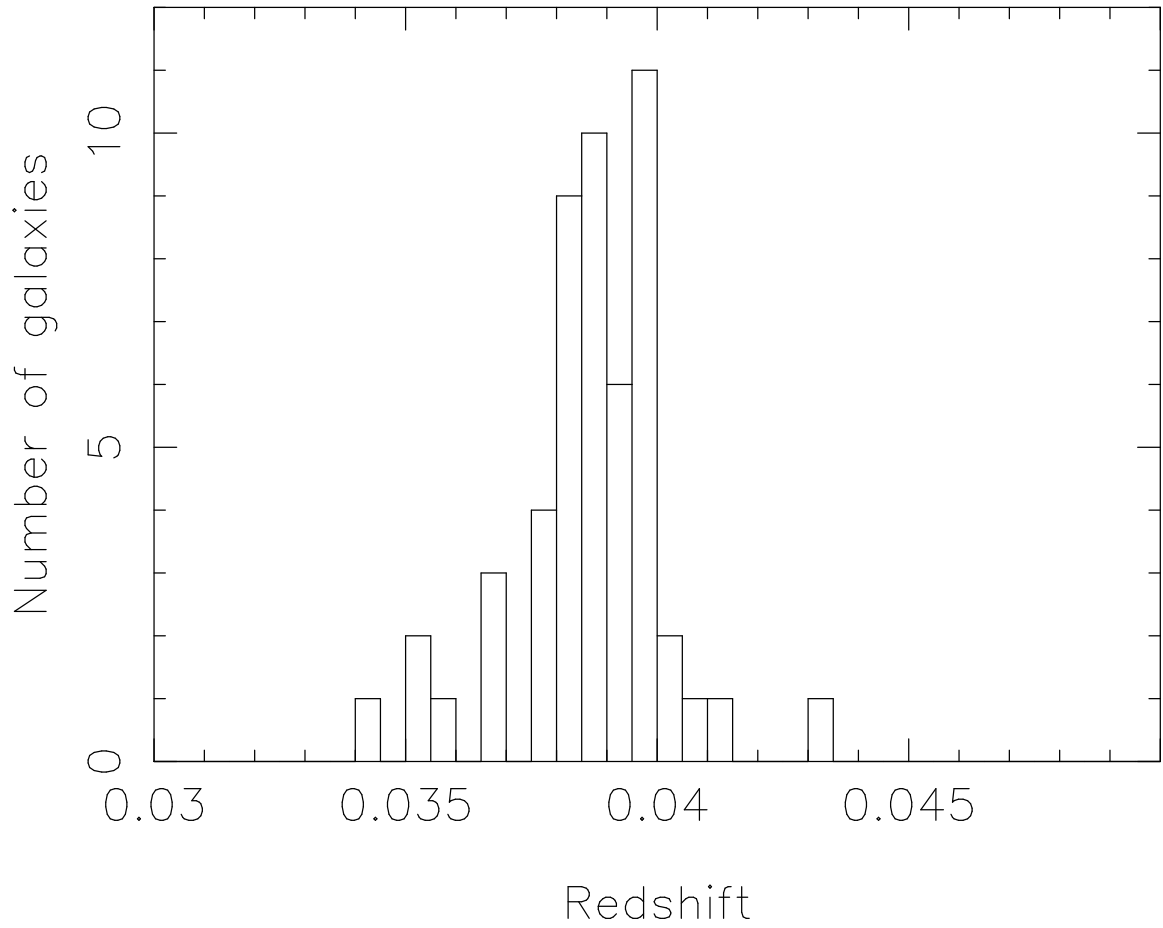}
\caption{
 Histogram of the redshift distribution of the galaxies within 
$1\degr$ of the host on the sky, and within the redshift range 
$z=0.03$--0.05.
 \label{fig8}
}
\end{figure}

\clearpage

\begin{figure}
\epsscale{1.0}
\plotone{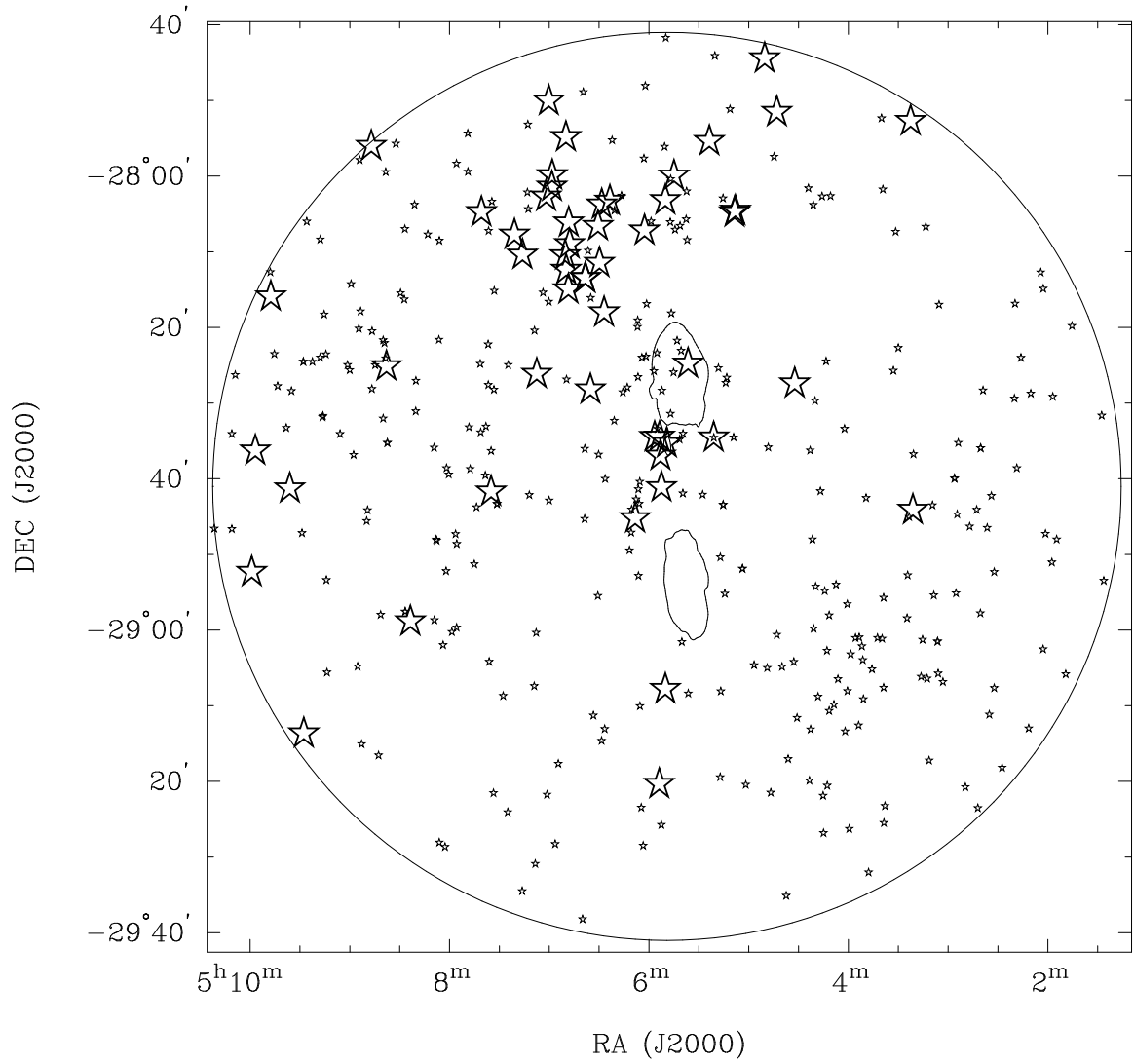}
\caption{ 
 The distribution of galaxies close to MSH~05$-$2{\it 2}; a 
$20\arcmin$ angular scale corresponds to a linear size of 0.9~Mpc. 
Galaxies with redshifts in the range $z=0.03$--0.05 are displayed. 
Large star symbols are for galaxies within $\pm 0.003$ of the host 
redshift at $z=0.038286$; galaxies outside this range are shown with 
small star symbols.  The large circle shows the $1\degr$ radius of 
the 2dF instrument.  A contour at 10~mJy~beam$^{-1}$ shows the lobes 
of the giant radio source at 843~MHz.
 \label{fig9}
}
\end{figure}

\clearpage

\begin{figure}
\epsscale{0.8}
\plotone{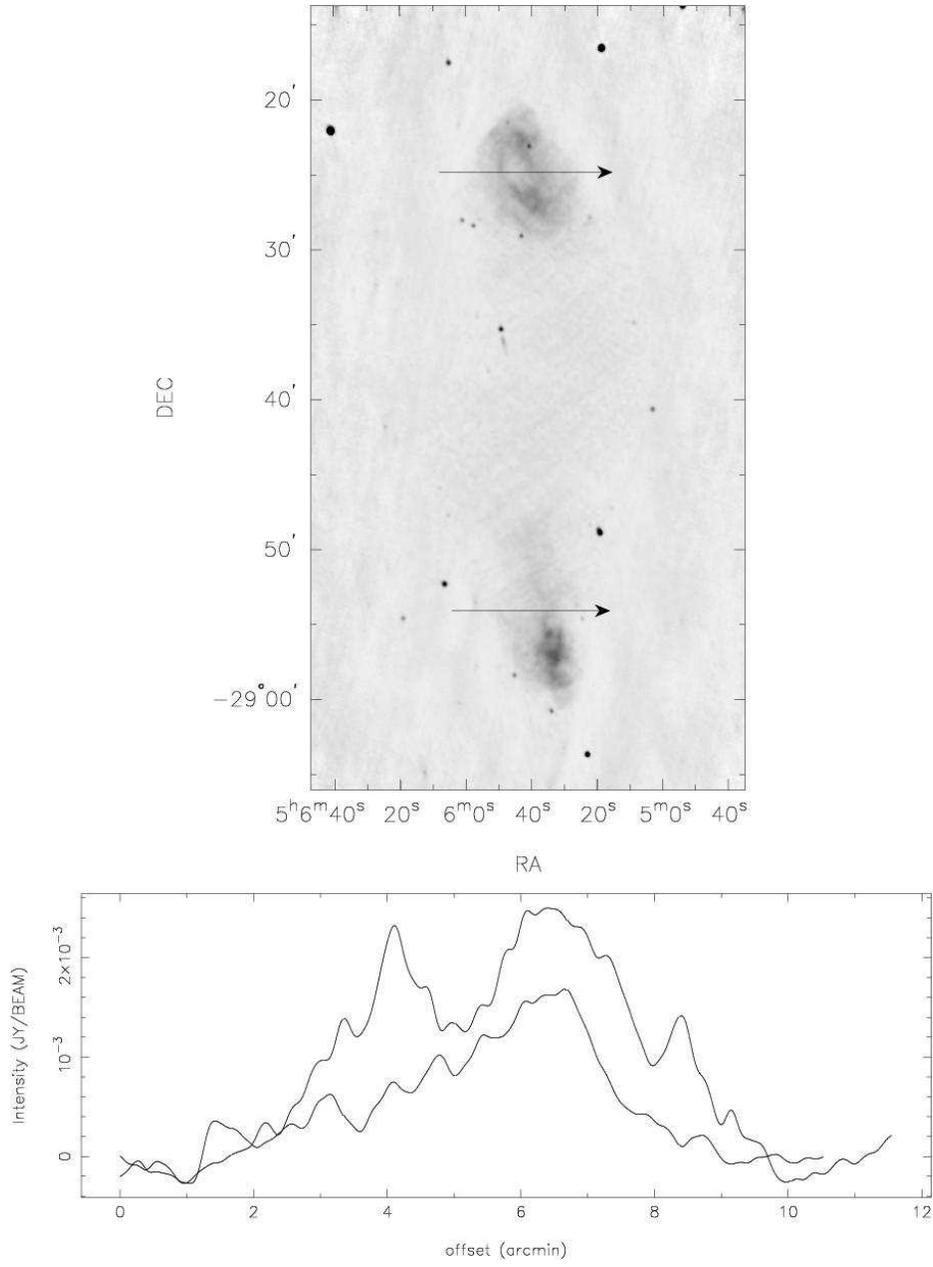}
\caption{Slice profiles across the two lobes of MSH~05$-$2{\it 2}.  
The radio image has a beam of FWHM $15\farcs5 \times 14\farcs2$ at a P.A. of $8\fdg6$.
\label{fig10}
}
\end{figure}

\clearpage

\begin{figure}
\epsscale{0.8}
\plotone{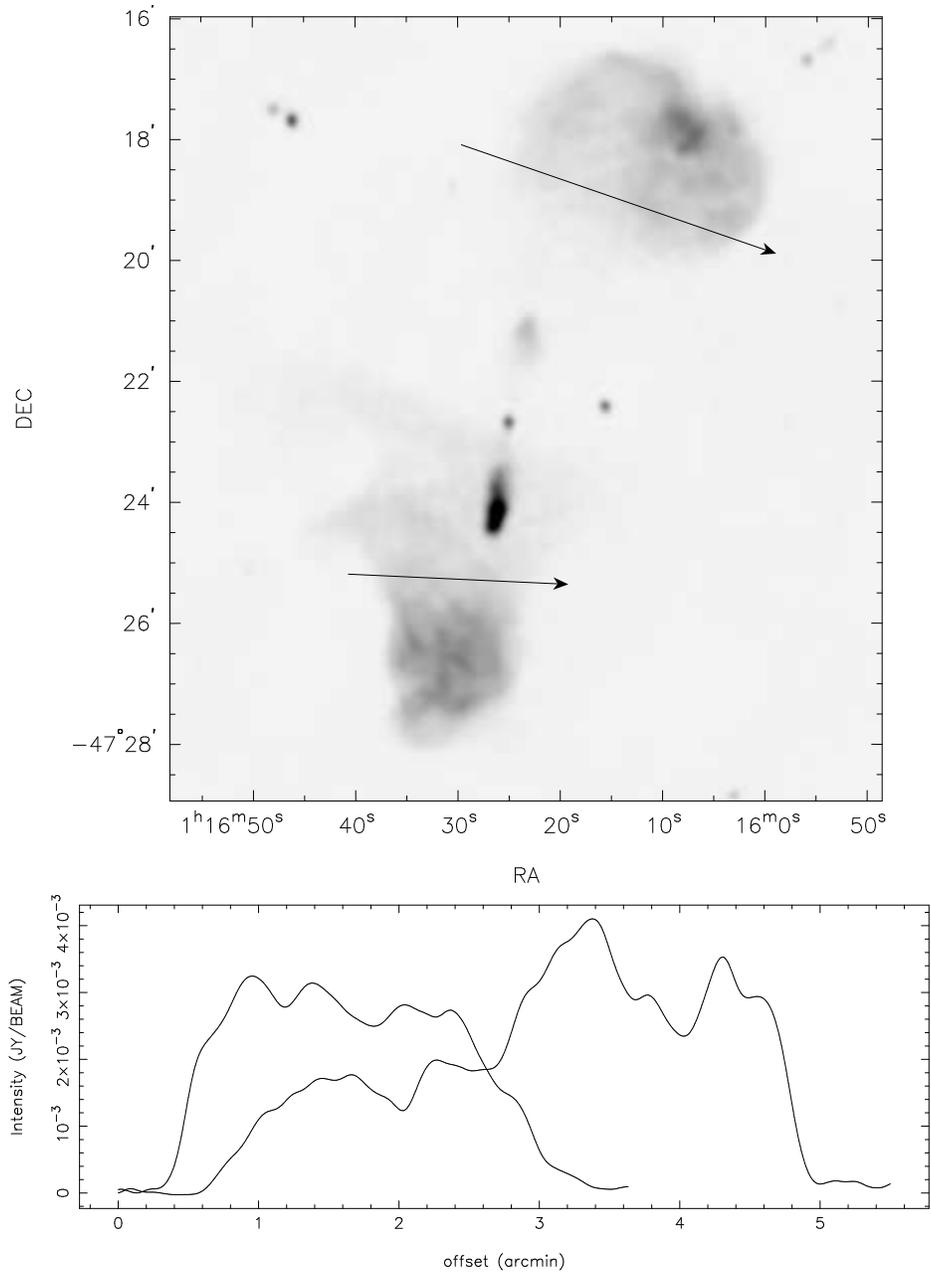}
\caption{Slice profiles across the two lobes of PKS~J0116$-$4722.  
The radio image has a beam of FWHM $10\farcs24 \times 9\farcs10$ at a P.A. of $10\fdg7$.
\label{fig11}
}
\end{figure}

\clearpage

\begin{figure}
\epsscale{0.8}
\plotone{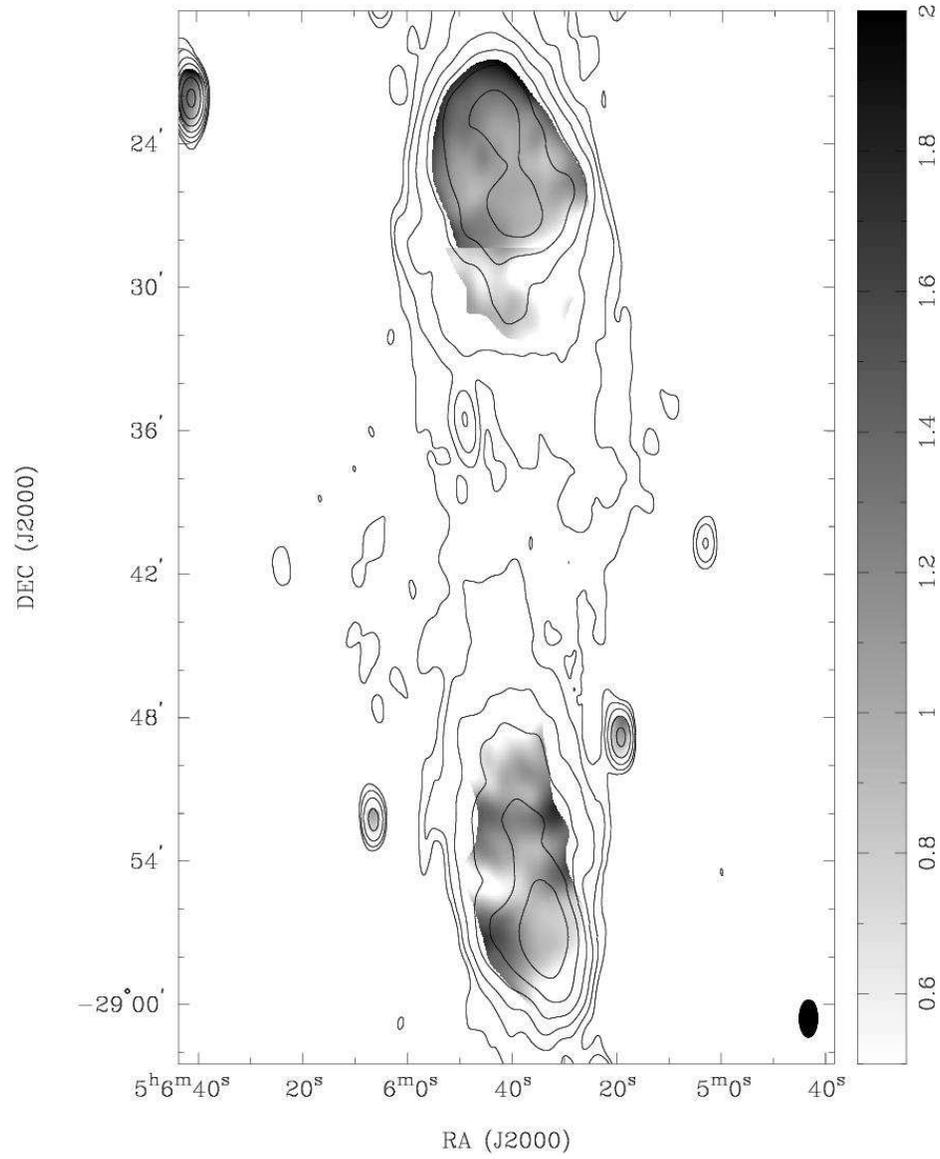}
\caption{Spectral index distribution between 843 and 1520~MHz, made using
  images with beam FWHM $94\farcs0 \times 45\farcs0$. Grey scales span the
  range $\alpha = -0.5$ to $-2.0$.
\label{fig12}
}
\end{figure}

\clearpage

\begin{figure}
\epsscale{0.8}
\plotone{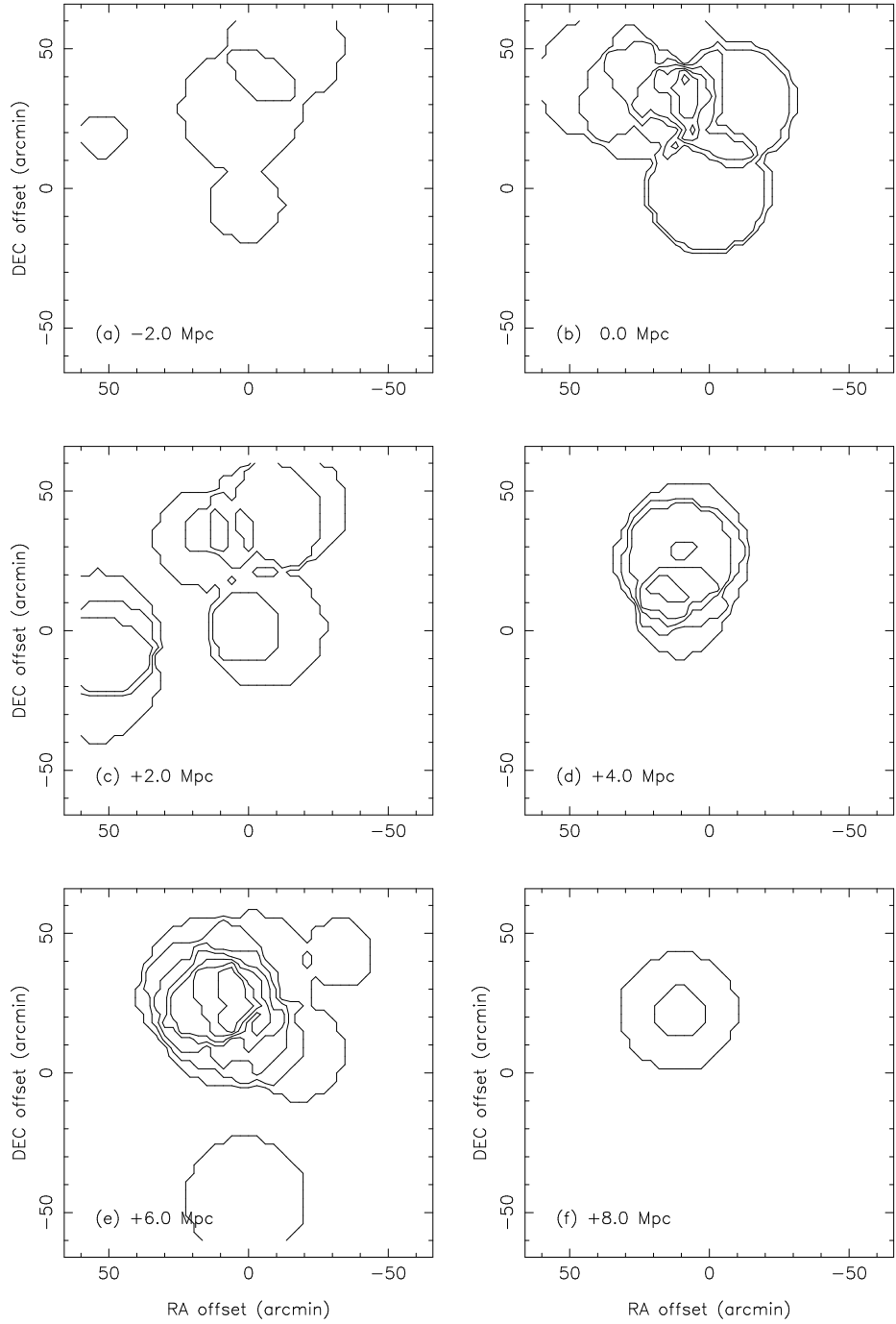}
\caption{Fractional overdensity distribution in the vicinity of
MSH~05$-$2{\it 2}. $50\arcmin$ angular scale corresponds to a linear size of
2.25~Mpc. Contours at $\Delta n / \overline{n} = $0, 2, 4, 6, 8, 10, 
and
12. A spherical top-hat smoothing function with radius $R=1$~Mpc was used.
\label{fig13}
}
\end{figure}

\end{document}